\documentclass[12pt]{article}
\usepackage[square]{natbib} 
\usepackage{a4}
\usepackage{latexsym} 
\usepackage{amssymb}  
\usepackage{graphicx}
\usepackage{epsfig}

\newcounter{zyxabstract}     
\newcounter{zyxrefers}        

\newcommand{\newabstract}
{\newpage\stepcounter{zyxabstract}\setcounter{equation}{0}
\setcounter{footnote}{0}}

\newcommand{\rlabel}[1]{\label{zyx\arabic{zyxabstract}#1}}
\newcommand{\rref}[1]{\ref{zyx\arabic{zyxabstract}#1}}

{\section*{References}\setcounter{zyxrefers}{0}
\begin{list}
{[\arabic{zyxrefers}]}
{\usecounter{zyxrefers}\setlength{\parindent}{0cm}\setlength{\itemsep}{0cm}}

}
{\end{list}}
\newenvironment{thebibliographynotitle}[1] 
{\section*{References}\setcounter{zyxrefers}{0}
\begin{list}{[\arabic{zyxrefers}]}
{\usecounter{zyxrefers}\setlength{\parindent}{0cm}\setlength{\itemsep}{-1.5mm}}}
{\end{list}}

\renewcommand{\bibitem}[1]{\item\rlabel{y#1}}
\renewcommand{\cite}[1]{[\rref{y#1}]}      
\newcommand{\citetwo}[2]{[\rref{y#1},\rref{y#2}]}
\newcommand{\citethree}[3]{[\rref{y#1},\rref{y#2},\rref{y#3}]}
\newcommand{\citefour}[4]{[\rref{y#1},\rref{y#2},\rref{y#3},\rref{y#4}]}

\newcommand{\vs}{}
\begin{document}
\begin{titlepage}
\begin{flushright} 
{\small
}
\end{flushright}

\begin{center}
{\LARGE\bf Strong interactions:\\[2.1mm]From methods to structures$^{*}$}
\\[1cm]
474. WE-Heraeus-Seminar\\
Physikzentrum Bad Honnef, Bad Honnef, Germany\\
February 12 --- 16, 2011\\[1cm]
{\bf Nora Brambilla}$^{1}$,  {\bf Evgeny Epelbaum}$^{2}$,
{\bf H.-W.~Hammer}$^3$ and {\bf Ulf-G.~Mei{\ss}ner}$^{3,4}$ 
\\[0.3cm]
$^1$Physik-Department T30f, Technische Universit\"at M\"unchen,\\
James-Frank-Str. 1, 85747 Garching, Germany\\[0.3cm]
$^2$Institut f\"ur Theoretische Physik II, Ruhr-Universit\"at Bochum,\\
44780 Bochum, Germany\\[0.3cm]
$^3$Helmholtz-Institut f\"ur Strahlen- und Kernphysik (Theorie) \\
and Bethe Center for Theoretical Physics, Universit\"at Bonn, \\ D-53115 Bonn,
Germany\\[0.3cm]
$^4$Institut f\"ur Kernphysik (Theorie), Institute for Advanced Simulation, \\
and J\"ulich Center for Hadron Physics, Forschungszentrum J\"ulich, \\ D-52425
J\"ulich, 
Germany\\[1cm]
{\sl Dedicated to the memory of our friend and colleague Klaus Goeke}\\[0.75cm]
{\large ABSTRACT}
\end{center}
These are the proceedings of the workshop on ``Strong interactions: From
methods to structures''
held at the Physikzentrum Bad Honnef of the
Deutsche Physika\-lische Gesellschaft, Bad Honnef, Germany from 
February 12 to 16, 2011. The workshop concentrated on physics of cold atoms,
chiral perturbation theory for mesons and baryons, chiral dynamics in
few-baryon systems and effective field theories for systems with heavy quarks. 
Included are a short contribution per talk.

\noindent\rule{6cm}{0.3pt}\\
\footnotesize{$^*$ This workshop was funded by the WE-Heraeus foundation.
This
research is part of the EU Research Infrastructure Integrating Activity 
``Study of Strongly Interacting Matter''
(acronym HadronPhysics2). 
}
\end{titlepage}

\section{Introduction}

Effective field theory techniques have become widely adopted 
in various fields of theoretical atomic, nuclear and particle physics. The main 
goal of this workshop was bringing together people using this 
methodology for studying different facets of strongly interacting 
matter to present the ongoing developments in various fields and 
foster discussions and exchange of knowledge. More specifically, 
the workshop focused on cold-atom physics, chiral perturbation theory and its
extensions, nuclear effective field theory and heavy quark physics. 
Some recent developments in lattice QCD have also been discussed in several
talks. 

This meeting followed the series of workshops in Ringberg (Germany), 1988,
Dobog\'ok\"o (Hungary), 1991, Karreb\ae ksminde (Denmark), 1993,
Trento (Italy), 1996 and Bad Honnef (Germany), 1998, 2001, 2004 and 2006.
All these workshops shared similar features,
about 50 participants, a fairly large amount of time devoted to discussions
rather than presentations and an intimate environment with lots of
discussion opportunities.

This meeting took place in February 2011 in the Physikzentrum Bad Honnef
in Bad Honnef, Germany, and the financial support provided by the 
WE--Heraeus--Stiftung allowed us to cover the local
expenses for all participants and to support the travel of a fair amount of
participants. The WE-Heraeus foundation also provided
the administrative support for the workshop in the person of the
able secretary Mrs.~Elisabeth Nowotka. We extend our sincere gratitude to the
WE-Heraeus Stiftung for the support and to Mrs.~Nowotka for the
precious help in the organization. We would also like to thank the staff of the
Physikzentrum for the excellent service given to us during the workshop
and last but not least the participants for making this an exciting
and lively meeting.

The meeting had 72 participants whose names, institutes and email addresses
are listed below. 47 of them presented results in talks of 35 minutes in
length. A short description of their contents and a list of the most relevant
references can be found below. As in the previous workshops we
felt that such mini-proceedings represent a more appropriate framework than
full-fledged proceedings. 
Most results are or will soon be published and available on the archive,
so this way we can achieve speedy publication and avoid duplication of
results in the archive.

Below follows first the program, then the list of participants 
followed by the abstracts of (most of) the talks which can also be
obtained from the workshop website 

\bigskip
\noindent
{\small \tt http://www.tp2.ruhr-uni-bochum.de/forschung/vortraege/workshops/bh11/}

\vspace{1.0cm}

\hfill N.~Brambilla, E.~Epelbaum, H.-W.~Hammer and U.-G.~Mei{\ss}ner

\newpage

\section{Program}
\begin{tabbing}
xx:xx \= A very very very long name \= \kill
{\bf \large Saturday, February 12th 2011}\\[4pt]
14:20 \>    Evgeny Epelbaum    \>  Introductory Remarks \\
\> (Bochum) \\ [6pt]
{\it  Early Afternoon  Session}\\
Chair: H.-W.~Hammer    \>  \> {\bf  Cold Atoms I} \\
  14:30 \>    Matt Wingate (Cambridge)     \>  
Monte Carlo computations for Fermi gases \\
15:05 \>      Lucas Platter (Chalmers) \> 
  How a few affect many\\
 15:40 \>
        Doerte Blume (Pullman)\>
        Universal features of weakly and strongly interacting \\
  \>\> few-body systems \\
16:15 \>\>{\em  Coffee}\\[6pt]
{\it   Late Afternoon Session}\\
Chair: H.-W.~Hammer    \>  \> {\bf  Cold Atoms II} \\
17:00 \>
        Joe Carlson (LANL)\>   Recent progress in simulations of cold atoms      \\
17:35 \>
        Mike Birse (Manchester)  \>
       Functional RG for few-body systems\\
       
18:10 \>
        Stefan Fl\"orchinger (CERN) \> Functional renormalization and ultracold quantum gases\\
18:45\>\>{\em End of Session}\\
19:00\>\> {\em Dinner }\\[18pt]
{\bf \large  Sunday, February 13th, 2011}\\[4pt]
{\it  Early Morning Session}\\
        Chair: H.-W.~Hammer \>\>{\bf  Chiral Perturbation Theory I}\\
09:00 \>
        Hans Bijnens (Lund)\>  CHPT in new surroundings\\
09:35\>
        Andre Walker-Loud \>
        Computing nucleon magnetic moments and electric \\
\> (Berkeley) \> polarizabilities with lattice QCD in background electric fields \\
10:10 \>
        Sebastien Descotes-Genon \>    
Dispersive analysis of isospin breaking in Kl4 decays \\
\> (Paris) \> \\
10:45 \>\> {\em Coffee}\\[6pt]
{\em Late Morning Session}\\
        Chair: H.-W.~Hammer\>\>{\bf   Chiral Perturbation Theory II}\\
11:05\>
        Jose Pelaez (Madrid)  \>  
Quark mass and $N_c$ dependence of meson-meson
        scattering: \\
\> \> Phases and resonances from standard and unitarized CHPT \\
11:40\>
        Peter Bruns (Regensburg)  \>   
Coupled channel Bethe-Salpeter approach to pion-nucleon \\
\> \> scattering\\
12:15\>\>{\em End of Session}\\
12:30\>\>{\em Lunch}\\[6pt]
{\em Early Afternoon Session}\\
        Chair: Nora Brambilla \>\>{\bf   Heavy Quark Physics I}\\
14:00\>
        Bernd Kniehl (Hamburg)    \>  Reconciling J/Psi production at HERA,
        RHIC, Tevatron and \\
\> \> LHC with NRQCD factorization at next-to-leading order \\
14:35\>
        Andre Hoang (Vienna)  \> Top quark mass at LHC \\
15:10\>
        Alex Rothkopf (Tokyo)   \> Proper heavy quark potential from lattice QCD \\
15:45\>
        Antonio Vairo (Munich)   \> The correlator of Polyakov loops at NNLO \\
16:20\>\>{\em Coffee}\\[6pt]
{\em Late Afternoon Session}\\
        Chair: Nora Brambilla \>\>{\bf Heavy Quark Physics II  }\\
17:00\>
        York Schroeder (Bielefeld)  \>   Quark mass effects in QCD thermodynamics \\
17:35\>
        Christoph Hanhart (J\"ulich)   \> How to identify hadronic molecules\\
18:10\>
        Peter Petreczky (BNL)  \>       Effective field theory approach for
        quarkonium at finite\\
\> \>   temperature\\
18:45\>\>        
{\em End of Session}\\
19:00\>\>        
{\em Dinner: Invitation by the Wilhelm und Else Heraeus-Stiftung }\\[16pt]
{\bf \large Monday, February 14th, 2011}\\[2pt]
{\em Early Morning Session}\\
        Chair: H.-W.~Hammer  \> \> {\bf Cold Atoms III}\\
09:00\>
        Yvan Castin (Paris)  \>  Four-body Efimov effect\\
09:35\>
        Arnoldas Deltuva (Lisbon)   \>  Universality in four-body scattering \\
10:10\>
        Dmitri Fedorov (Aarhus) \> Three boson systems near a Feshbach resonance in a \\
\>\>two-channel zero-range model\\
10:45\>\>        
{\em Coffee}\\[6pt]
{\em Late Morning Session}\\
        Chair: H.-W.~Hammer  \>\>{\bf  Cold Atoms IV}\\
11:05\>
        Kerstin Helfrich (Bonn) \> Three bosons in two dimensions \\
11:40\>
        Dima Petrov (Paris) \>  Parametric excitation of a 1D gas in
        integrable and \\
\> \> non-integrable cases \\
12:15\>\>{\em End of Session}\\
12:30\>\>{\em Lunch}\\[6pt]
{\em Early Afternoon Session}\\
        Chair: Ulf-G.~Mei{\ss}ner  \> \>  {\bf Chiral Perturbation Theory III}\\
14:00\>
        Akaki Rusetsky (Bonn)  \>   Effective field theories in a finite volume \\
14:35\>
        Feng-Kun Guo (Bonn) \> Extraction of light quark mass ratio from heavy
        quarkonia \\
\> \> transitions  \\
15:10\>
        Bastian Kubis (Bonn)  \>   Rescattering effects in $\eta$ and $\eta '$ decays \\
15:45\>
        Norbert Kaiser (Munich)  \>  Low-energy pion-photon reactions and chiral symmetry  \\
16:20\>\>        
{\em Coffee}\\[6pt]
{\em Late Afternoon Session}\\
        Chair: Evgeny Epelbaum    \> \>{\bf Few-Baryon Systems I}\\
17:00\>
        Ruprecht Machleidt \>     The nuclear force problem: have we finally
        reached  the end \\
\> (Moscow, Idaho) \>of the tunnel? \\
17:35\>
        Hermann Krebs (Bochum)   \>  Three-nucleon forces with explicit $\Delta$ fields \\
18:10\>
        Nasser Kalantar-Nayesta-    \>  What have we learned about
        three-nucleon forces at \\  
\> naki (KVI)\> intermediate energies \\
18:45\> \>{\em End of Session}\\
19:00\> \>{\em Dinner}\\[18pt]
{\bf \large Tuesday, February 15th, 2011}\\[4pt]
{\em Early Morning Session}\\
        Chair: Evgeny Epelbaum   \> \>{\bf  Few-Baryon Systems II}\\
09:35\>
        Silas Beane     \>    Hadron interactions from
        lattice QCD\\  
\>(New Hampshire/Bern) \> \\
10:10\>
        Dean Lee  (North Carolina)  \>  Nuclear physics from lattice effective
        field theory  \\
10:45\>\>{\em Coffee}\\[6pt]
{\em Late Morning Session}\\
        Chair: Ulf-G. Mei\ss ner\>  \>{\bf  Hadronic contributions}\\
11:05\>
        Hartmut Wittig (Mainz)    \>   The hadronic vacuum polarization
        contributions to $(g-2)_\mu$ \\
\> \> from lattice QCD  \\
11:40\>
        Antonio Pineda  \>  The muonic hydrogen Lamb shift and the proton
        radius \\
\> (Barcelona) \\
12:15\> \>{\em End of Session}\\
12:30\> \>{\em Lunch}\\[6pt]
{\em Early Afternoon Session}\\
        Chair: Nora Brambilla   \> \> {\bf Heavy Quark Physics III}\\
14:00\>
        Binsong Zou (IHEP)  \>  Prediction of super-heavy $N^\star$ and
        $\Lambda^\star$ with
        hidden charm   \\
\> \> and beauty \\
14:35\>
        Qiang Zhao (IHEP)     \>    A coherent view of the charmonium hadronic
        and radiative  \\ 
\> \> decays \\
15:10\>
        Mei Huang (IHEP)    \>   Interplay between chiral and deconfinement
        phase transitions   \\ 
15:45\> \>        {\em Coffee}\\ [6pt]
{\em Late Afternoon Session}\\
        Chair: Nora Brambilla \> \> {\bf  Heavy Quark Physics IV}\\
17:00\>
        Jacopo Ghiglieri (Munich)   \>  Effective field theories for heavy
        quark(onia) at finite  \\ 
\>\> temperature \\
17:35\>
        Paul Romatschke   \>  Hadrodynamics: a tool for strongly
        coupled systems   \\
\> (Frankfurt) \\
18:10\>
        Yu Jia (IHEP)  \>  Some novel development in quarkonium
        electromagnetic \\
\> \> transitions \\ 
18:45\> \>{\em End of Session}\\
19:00\> \>{\em Dinner}\\[18pt] \\ \\
{\bf \large Wednesday, February 16th, 2011}\\[4pt]
{\em Early Morning Session}\\
        Chair: Evgeny Epelbaum \> \>  {\bf Few-Baryon Systems III}\\
09:00\>
        Daniel Phillips   \>  Electromagnetic properties of
        single-neutron halo nuclei    \\
 \>(Athens, Ohio) \> from effective
  field theory \\
09:35\>
        Rocco Schiavilla (JLab) \>  Electromagnetic processes on few-nucleon
        systems at low \\
\> \> energies \\
10:10\>
        Doron Gazit (Jerusalem) \>  Inferring nuclear structure from
        electroweak reactions  \\
10:45\> \>{\em Coffee}\\[6pt]
{\em Late Morning Session}\\
        Chair: Evgeny Epelbaum \> \> {\bf Few-Baryon Systems IV }\\
11:05\>
        Vadim Baru (J\"ulich))  \>   A high-accuracy calculation of the
        pion-deuteron scattering \\
\> \> length  \\
11:40\>
        Johan Haidenbauer  \>  Hyperon-nucleon and hyperon-hyperon
        interactions in chiral   \\
\> (J\"ulich) \> effective field theory \\
12:15\> \>{\em End of Session}\\
12:30\> \>{\em Lunch}\\[6pt]
{\em Early Afternoon Session}\\
        Chair: Nora Brambilla   \> \> {\bf Heavy-Quark Physics V}\\
14:00\>  Jianwei Qiu (BNL)  \> QCD factorization and heavy quarkonium production    \\
14:35\>  Pietro Falgari (Utrecht) \>Threshold resummation of heavy coloured
particle cross \\
\> \> section   \\
15:10 \>  Evgeny Epelbaum  \> Concluding remarks\\
\> (Bochum) \\
15:20\> \>{\em End of Workshop}\\
\end{tabbing}

\section{Participants and their email}

\begin{tabbing}
A very long namexxxxxxx\=a very long institutexxxxxxxxx\=email\kill
V. Baru \> FZ J\"ulich \> v.baru@fz-juelich.de \\
S.\,R. Beane \> Univ. Bern/New Hampshire \> silas@physics.unh.edu\\
M. Berwein \> TU M\"unchen \> matthias.berwein@mytum.de \\
J. Bijnens \> Lund Univ. \> bijnens@thep.lu.se \\
M. Birse \> Univ. of Manchester \> mike.birse@manchester.ac.uk \\
D. Blume \> Washington State Univ. \> doerte@wsu.edu \\
N. Brambilla \> TU M\"unchen \> Nora.Brambilla@ph.tum.de\\
P. Bruns \> Univ.\ Regensburg\>Peter.Bruns@physik.uni-regensburg.de\\
J. Carlson \> LANL \> carlson@lanl.gov \\
Y. Castin \> Lab. Kastler Brossel \> yvan.castin@lkb.ens.fr \\
M. Cleven \> FZ J\"ulich \> m.cleven@fz-juelich.de\\
A. Deltuva \> Univ. of Lisbon \> deltuva@cii.fc.ul.pt\\
S. Descotes-Genon \>CNRS  \> sebastian.descotes@th.u-psud.fr\\
C. Ditsche \> Univ. Bonn \> ditsche@hiskp.uni-bonn.de \\
E. Epelbaum \> Ruhr-Univ. Bochum \> evgeny.epelbaum@rub.de\\
P. Falgari \> Univ. Utrecht \> p.falgari@uu.nl\\
D. Fedorov \> Aarhus Univ. \> fedorov@phys.au.dk\\
A. Filin \> Univ. Bonn \> a.filin@fz-juelich.de \\
S. Fl\"orchinger \> CERN \> stefan.floerchinger@cern.ch \\
D. Gazit \> Hebrew Univ. of Jerusalem \> doron.gazit@mail.huji.ac.il \\
J. Ghiglieri \> TU M\"unchen \> jacopo.ghiglieri@ph.tum.de \\
F.-K. Guo \> Univ. Bonn \> fkguo@hiskp.uni-bonn.de\\
J. Haidenbauer \> FZ J\"ulich \> j.haidenbauer@fz-juelich.de\\
H.-W. Hammer \> Univ. Bonn \> hammer@hiskp.uni-bonn.de\\
C. Hanhart \> FZ J\"ulich \> c.hanhart@fz-juelich.de \\
K. Helfrich \> Univ. Bonn \> helfrich@hiskp.uni-bonn.de \\
A. Hoang \> Univ. Wien \> ahoang@mpp.mpg.de \\
M. Huang \> IHEP Beijing \> huangm@ihep.ac.cn \\
Y. Jia \> IHEP Beijing \> jiay@mail.ihep.ac.cn\\
N. Kaiser \>TU M\"unchen\> nkaiser@ph.tum.de \\
N. Kalantar-Nayestanaki \> KVI \> nasser@kvi.nl \\
B. Kniehl \> Univ. Hamburg \> kniehl@desy.de \\
S. K\"olling \> FZ J\"ulich \> s.koelling@fz-juelich.de\\
S. K\"onig \> Univ. Bonn \> sekoenig@hiskp.uni-bonn.de\\
J. Koschinski \> Ruhr-Univ. Bochum \> julia.koschinski@tp2.rub.de \\
H. Krebs \> Ruhr-Univ. Bochum \>  Hermann.Krebs@tp2.rub.de\\
B. Kubis \> Univ.\ Bonn \> kubis@hiskp.uni-bonn.de\\
B. Kwapinski \> Ruhr-Univ. Bochum \> Bartholomaeus.Kwapinski@tp2.rub.de\\
D. Lee \> NC State Univ. \> djlee3@unity.ncsu.edu\\
X.-H. Liu \> IHEP Beijing \> xhliu@ihep.ac.cn\\
I. Lorenz \> Univ. Bonn \> lorenzi@hiskp.uni-bonn.de\\
R. Machleidt \> Univ. of Idaho \> machleid@uidaho.edu \\
M. Mai \> Univ. Bonn \> mai@hiskp.uni-bonn.de\\
H. Martinez \> TU M\"unchen \> hector.martinez@ph.tum.de\\
U.-G. Mei{\ss}ner \> Univ.\ Bonn \& FZ J\"ulich \> meissner@hiskp.uni-bonn.de\\
D. Minossi \> FZ J\"ulich \> minossi@fz-juelich.de\\
J.R. Pelaez \> Univ. Madrid \> jrpelaez@fis.ucm.es\\
P. Petreczky \> BNL \> petreczk@quark.phy.bnl.gov\\
D. Pertov \> Univ. Paris-Sud \> dmitry.petrov@u-psud.fr \\
D. Phillips \> Ohio Univ. \> phillips@phy.ohiou.edu \\
P. Pietrulewicz \> TU M\"unchen \> Piotr.Pietrulewicz@ph.tum.de \\
A. Pineda \> Univ. Barcelona \> pineda@ifae.es\\
L. Platter \>Chalmers Univ. \> platter@chalmers.se\\
J.-W. Qiu \> BNL/SBU \> jwq@iastate.edu\\
P. Romatschke \> Univ. Frankfurt \> romatschke@fias.uni-frankfurt.de \\
A.K. Rothkopf \> Univ. of Tokyo \> rothkopf@nt.phys.s.u-tokyo.ac.jp\\
A. Rusetsky \>  Univ.\ Bonn \> rusetsky@itkp.uni-bonn.de\\
R. Schiavilla \> Old Dominion Univ./Jlab \> schiavil@jlab.org     \\
S. Schneider \> Univ. Bonn \> schneider@hiskp.uni-bonn.de \\
Y. Schr\"oder \> Univ. Bielefeld \> yorks@physik.uni-bielefeld.de\\
J. Sieverding \> Ruhr-Univ. Bochum \> jan.sieverding@tp2.rub.de\\
F. Stollenwerk \> FZ J\"ulich \> f.stollenwerk@fz-juelich.de\\
P. Sturm \> Ruhr-Univ. Bochum \> Patrick.Sturm@tp2.rub.de\\
S. T\"olle \> Univ. Bonn \> toelle@hiskp.uni-bonn.de\\
A. Vairo\> TU M\"unchen \> antonio.vairo@ph.tum.de\\
A. Walker-Loud \> Berkeley National Lab \> awalker-loud@lbl.gov\\
E. Wilbring \> Univ. Bonn \> wilbring@hiskp.uni-bonn.de\\
M. Wingate \> Univ. of Cambridge \> M.Wingate@damtp.cam.ac.uk \\
H. Wittig \> Univ. Mainz \> wittig@kph.uni-mainz.de \\
T. Wolf \> Ruhr-Univ. Bochum \> wolf.till@googlemail.com \\
Q. Zhao \> IHEP Beijing \> zhaoq@ihep.ac.cn \\
B. Zou \> IHEP Beijing \> zoubs@ihep.ac.cn \\
\end{tabbing}

\newabstract 
\begin{center}
{\large\bf Monte Carlo Calculations for Fermi Gases}\\[0.5cm]
Olga Goulko and {\bf Matthew Wingate}\\[0.3cm]
DAMTP, University of Cambridge, Wilberforce Road\\
Cambridge CB3 0WA, United Kingdom\\[0.3cm]
\end{center}

Ultracold atomic gases have become a fertile environment for cultivating and
examining diverse varieties of physical phenomena.  Many theories of condensed
matter can be realized through inventive trapping techniques.  Furthermore,
the purity of the gases permits accurate, systematically improvable
calculations, often employing field theoretic methods.

Our work has focused on Monte Carlo calculations for Fermi gases \cite{GPS},
especially in the unitary scattering limit.  Here, there are dimensionless
numbers and functions which are universal, depending only on one
characteristic length scale, say the mean interparticle spacing.  One role of
numerical calculations is to determine these universal properties from first
principles.

A number of Monte Carlo methods have been applied to this problem \cite{DLee}.
In our recent and ongoing work studying the properties of the unitary Fermi
gas near the superfluid/normal critical temperature, we employ a
generalization of diagrammatic determinant Monte Carlo \cite{BPST}.  We found
a reduction in autocorrelation time with a new type of configuration update.
We also explored the slightly spin-imbalanced unitary Fermi gas using a
sign-quenched method of updating, where the effects of the sign are taken into
account in observables.  For the critical temperature we find \cite{GW_pra}
\[
  T_c(\Delta\mu)/\varepsilon_F ~=~ 0.171(5) 
  \;+\; T_2\,(\Delta \mu/\varepsilon_F)^2 \,.
\]
For our data $T_2$ is consistent with 0, and from the fit uncertainty we
estimate a lower bound $T_2 > -0.5$ (at $68\%$ CL).  We find for the chemical
potential $\mu/\varepsilon_F = 0.429(7)$, independent of $\Delta\mu$ within
uncertainties for $\Delta\mu/\varepsilon_F < 0.2$.

Another quantity we can compute is Tan's contact.  This enters as the
sole long-distance quantity in OPE relations \cite{Platter}.
For the spin-balanced gas at $T=T_c$, our preliminary result for the contact
is \cite{GW_lattice2010}
\[
C(T_c) ~=~ 0.1102(11)\,\varepsilon_F^2 V ~=~ 3.26(3) \,k_F N \,.
\]

\setlength{\bibsep}{0.0em}
\begin{thebibliographynotitle}{99}
\bibitem{GPS} S.~Giorgini, L.P.~Pitaevskii, S.~Stringari, Rev.\ Mod.\
  Phys.\ \textbf{80} (2008) 1215.
\bibitem{DLee} D.~Lee, Prog.\ Part.\ Nucl.\ Phys.\ \textbf{63} (2009) 117.
\bibitem{BPST} E.~Burovski \textit{et al.}, New J.~Phys.\ \textbf{8} (2006) 
  153.
\bibitem{GW_pra} O.~Goulko and M.~Wingate, Phys.\ Rev.\ A \textbf{83} (2010) 
  053621.
\bibitem{Platter} L.~Platter, talk at this Seminar.
\bibitem{GW_lattice2010} O.~Goulko and M.~Wingate, PoS (Lattice 2010) 187,
  arXiv:1011.0312.
\end{thebibliographynotitle}

\newabstract 
\begin{center}
{\large\bf From Few to Many}\\[0.5cm]
{\bf Lucas Platter}\\
Fundamental Physics,\\
Chalmers University of Technology,\\
SE-41133 G\"oteborg, Sweden\\[0.3cm]
\end{center}
Strongly interacting many-body systems pose a constant challenge to
theory. For example, most results for non-relativistic spin-1/2
particles interacting through a large scattering length have been
obtained numerically~\cite{Bulgac:2010dg}. However, Shina Tan recently
derived analytically and with novel methods remarkable universal relations
that hold for any state of this system \cite{Tan123}. They are particularly
important since the central quantity (the so-called contact) controls
amongst other things the thermodynamics of the many-body ensemble.
Several other methods have been used to rederive these relations (see
e.g. Ref.~\cite{Braaten:2010if} for a summary) but one particular
powerful one was discussed in this talk. In Ref.~\cite{Braaten:2008uh}
it was shown that these relations can be rederived using the so-called
operator product expansion which is a short-distance
expansion that has been employed to various strongly-interacting
systems (e.g. ising model). This method has also
allowed to derive new universal relations for the fermionic
many-body system that were previously unknown. We have recently applied this approach also to
bosonic systems with a large scattering length
\cite{Braaten:2011sz} and have derived a number of new universal
relations. This system is special because it will display the Efimov
effect. This is encoded in the quantum field theoretical description
through a three-body force with an anomalous scaling
dimension~\cite{Bedaque:1998kg}. It's experimental signature is
discrete scale invariance and it is of importance in atomic, nuclear
and particle physics~\cite{Hammer:2010kp}. The versatility of the
operator product expansion guarantees that further universal relations
for additional systems (e.g. fermionic systems with three spin
states) can be derived. The importance of the three-body force and
therefore of the Efimov effect shows up as a second ``contact'' in the
bosonic universal relations. 

\setlength{\bibsep}{0.0em}
\begin{thebibliographynotitle}{99}
\bibitem{Bulgac:2010dg}
  A.~Bulgac, M.~M.~Forbes, P.~Magierski,
  arXiv:1008.3933 .

\bibitem{Tan123}
S.~Tan,
Annals of Physics {\bf 323}, 2952 (2008);
%
ibid.~{\bf 323}, 2971 (2008);
%
ibid.~{\bf 323}, 2987 (2008).

\bibitem{Braaten:2010if}
  E.~Braaten,
  [arXiv:1008.2922 [cond-mat.quant-gas]].

\bibitem{Braaten:2008uh}
  E.~Braaten, L.~Platter,
  Phys.\ Rev.\ Lett.\  {\bf 100}, 205301 (2008).

\bibitem{Braaten:2011sz}
  E.~Braaten, D.~Kang, L.~Platter,
  arXiv:1101.2854 .

\bibitem{Bedaque:1998kg}
  P.~F.~Bedaque, H.-W.~Hammer, U.~van Kolck,
  Phys.\ Rev.\ Lett.\  {\bf 82}, 463-467 (1999).

\bibitem{Hammer:2010kp}
  H.~-W.~Hammer, L.~Platter,
  Ann.\ Rev.\ Nucl.\ Part.\ Sci.\  {\bf 60}, 207-236 (2010).

\end{thebibliographynotitle}

\newabstract 
\begin{center}
{\large\bf Universal features of 
weakly and strongly interacting few-body systems}\\[0.5cm]
{\bf D. Blume}, K. M. Daily, and  D. Rakshit \\[0.3cm]
Department of Physics and Astronomy,
Washington State University,
  Pullman, Washington 99164-2814, USA\\[0.3cm]
\end{center}

Universality arises in many different branches
of physics.
Generally speaking, a given phenomenon is referred to as universal
when the observables for different physical 
systems can be collapsed to a single point or curve.
The talk given at the
``Strong interactions:
From methods to structures'' workshop  summarized 
our theoretical investigations of universal
aspects of ultracold bosonic and fermionic gaseous few-atom 
systems, which share certain features with 
few-nucleon systems and can thus be regarded as model systems.
In the ultracold regime, where the deBroglie
wave length is large compared to the range of the
atom-atom interaction potentials, the collisions between 
atoms become so slow that the details of the 
interactions are, to leading order, neglegible. 
In this regime, the dynamics of few-atom systems
is governed by a few ``effective parameters''
(such as the s-wave scattering length) and independent of 
the details of the underlying two-body potentials. 
As a consequence, the true atom-atom potential can be replaced by
a model potential that is most suitable for the numerical
or analytical techniques employed. 
For example, we employ a ``soft''
Gaussian potential in our stochastic variational
studies~\citethree{cgbook}{stec08}{dail10}
and a zero-range pseudo-potential in our analytical analysis.
As a first example, we analyzed the behavior of weakly-interacting 
Bose gases in terms of effective, and presumably universal, 
N-body interactions. This study is partially
motivated by Ref.~\cite{john09}. As a second example, we investigated the 
behavior of strongly-interacting two-component Fermi gases
with equal and unequal masses~\citetwo{blum10}{gand10}. 
Implications of our studies for on-going experimental 
efforts were discussed.
Support by the NSF through
grant PHY-0855332 and the ARO
is gratefully acknowledged.

\setlength{\bibsep}{0.0em}
\begin{thebibliographynotitle}{99}
\bibitem{cgbook}
Y.~Suzuki and K.~Varga.
\newblock {\em Stochastic Variational 
Approach to Quantum Mechanical Few-Body Problems}.
\newblock Springer Verlag, Berlin, 1998.
\bibitem{stec08}
J.~{von Stecher}, C.~H. Greene, and D.~Blume.
Phys. Rev. A77 (2008) 043619.
\bibitem{dail10}
K.~M. Daily and D.~Blume.
Phys. Rev. A81 (2010) 053615.
\bibitem{john09}
P.~R. Johnson {\it et al.,}
New J. Phys. 11 (2009) 093022.
\bibitem{blum10}
D.~Blume and K.~M. Daily.
Phys. Rev. Lett. 105 (2010) 170403.
Phys. Rev. A82 (2010) 063612.
\bibitem{gand10}
S.~Gandolfi and J.~Carlson,
arXiv:1006.5186.
\end{thebibliographynotitle}


\newabstract 
\begin{center}
{\large\bf Functional RG for few-body systems}\\[0.5cm]
{\bf Michael C. Birse }\\[0.3cm]
Theoretical Physics Group, School of Physics and Astronomy,\\
The University of Manchester, M13 9PL, UK\\[0.3cm]
\end{center}

The functional renormalisation group for the one-particle-irreducible effective 
action has proved to be a powerful tool for analysing a variety of physical
systems \cite{btw}. In particular, it has been applied to fermionic matter close 
to the unitary limit \citetwo{bkmw}{dgpw}. As a field-theoretic approach, it can be 
easily matched on to the effective field theories that are now being used to 
describe both nuclear and atomic systems at low energies.

The input into these many-body calculations is fixed by applying the same RG  
to few-body systems. The first studies used only two-body physics, within a local 
trucation of the action \citethree{bkmw}{dgpw}{b08}. This was then extended to include 
three-body interactions, again with a local form \citetwo{fsmw}{bwk10}. In three-body
systems where the particle exchange force is attractive, an Efimov limit cycle is 
obtained, with a very good approximation to the discrete scaling factor \citetwo{fsmw}{sm}.
The full set of local terms needed to describe the four-body system (dimer-dimer scattering) includes terms where one of the dimers can break up into its constituent 
atoms, as noted by Schmidt and Moroz for the case of bosons \cite{sm}. 

Four-fermion systems with a full local truncation of the action were studied in 
Ref.~\cite{bkw11}. The dimer-dimer scattering length obtained with the RG agrees
very well with the value from standard few-body calculations. Moreover the result
is only weakly dependent on the choice of the regulator function used in the RG evolution. Anomalous dimensions for three- and four-body forces have also been 
determined using this method and, in the fermionic case, these show that such 
forces are all highly irrelevant.

\setlength{\bibsep}{0.0em}
\begin{thebibliographynotitle}{99}
\bibitem{btw}J. Berges, N. Tetradis and C. Wetterich, Phys. Rept. {\bf 363} 
(2002) 223.
\bibitem{bkmw} M. C. Birse, B. Krippa, J. A. McGovern and N. R. Walet,
Phys. Lett. B \textbf{605} (2005) 287.
\bibitem{dgpw} S. Diehl, H. Gies, J. M. Pawlowski and C. Wetterich, 
Phys. Rev. A \textbf{76} (2007) 021602; 053627.
\bibitem{b08} M. C. Birse, Phys. Rev. C \textbf{77} (2008) 047001.
\bibitem{fsmw} S. Floerchinger,  R. Schmidt, S. Moroz and C. Wetterich, 
Phys. Rev. A \textbf{79} (2009) 013603; 042705.
\bibitem{bwk10} M. C. Birse, B. Krippa and N. R. Walet, 
Phys. Rev. A \textbf{81} (2010) 043628.
\bibitem{sm} R. Schmidt and S. Moroz, 
Phys. Rev. A \textbf{81} (2010) 052709. 
\bibitem{bkw11} M. C. Birse, B. Krippa and N. R. Walet, 
Phys. Rev A \textbf{83} (2011) 023621.
\end{thebibliographynotitle}

\newabstract 
\begin{center}
{\large\bf Functional renormalization and ultracold quantum gases}\\[0.5cm]
{\bf Stefan Floerchinger}\\[0.3cm]
Physics Department, Theory Unit, CERN\\
CH-1211 Gen\`eve 23, Switzerland\\[0.3cm]
\end{center}

Functional renormalization is a modern method mainly used to study strongly interacting field theories. An exact flow equation for the one-particle irreducible effective action \cite{Wetterich} is solved approximately by truncating the most general form of the action.

This method was employed to study different aspects of ultracold quantum gases. For example, a detailed picture of the BCS-BEC crossover for fermions with two components has been obtained \cite{FSW}. Observables such as the gap, the radius of the Fermi sphere or the dispersion relation have been calculated for all values of the crossover parameter $a k_F$ including the unitarity point where it diverges. The critical temperature interpolates between a weakly interacting Fermi gas described by BCS theory including Gorkov's correction and a gas of bosonic bound states with repulsive interaction.

For interacting bosons with local repulsive interaction many thermodynamic properties have been calculated in the superfluid phase. This includes the temperature and density dependence of pressure, energy and entropy-density, superfluid and condensate-fraction, correlation length, specific heat, isothermal and adiabatic compressibility and various sound velocities \cite{FWThermo}.

Functional renormalization was also used successfully to study interesting few-body physics (see the contribution of M. Birse to this conference). One example is the Efimov effect for three species of fermions with identical mass and scattering length. In the functional renormalization group description this shows up as a limit cycle \citetwo{FSMW}{MFSW}. Based on a continuity argument one can predict quantum phase transitions from a BCS type phase to one dominated by three-body bound states and further to a BEC type phase \cite{FSMW}. After extending the approach to non-equal scattering length it can be used to explain recently measures three-body loss rates in lithium \cite{FSWLithium}.

\setlength{\bibsep}{0.0em}
\begin{thebibliographynotitle}{99}
\bibitem{Wetterich} C. Wetterich, Phys. Lett. B 301 (1993) 90.
\bibitem{FSW} S. Floerchinger {\it et al.,}
Phys. Rev. A 81 (2010) 063619.
\bibitem{FWThermo} S. Floerchinger and C. Wetterich, Phys. Rev. A 79 (2009) 063602.
\bibitem{FSMW} S. Floerchinger {\it et al.,}
Phys. Rev. A 79 (2009) 013603.
\bibitem{MFSW}S. Moroz {\it et al.,}
Phys. Rev. A 79 (2009) 042705.
\bibitem{FSWLithium} S. Floerchinger {\it et al.,}
Phys. Rev. A 79 (2009) 053633.
\end{thebibliographynotitle}

\newabstract 
\begin{center}
{\large\bf Chiral Perturbation Theory in New Surroundings}\\[0.5cm]
{\bf Johan Bijnens}\\[0.3cm]
Department of Astronomy and Theoretical Physics\\
Lund University, S\"olvegatan 14A, SE22362 Lund, Sweden\\[0.3cm]
\end{center}

This talk describes some of the recent work done in Lund in Chiral
Perturbation Theory (ChPT) and Effective Field Theory (EFT).

It recently became clear that ChPT can also be applied to situations
where momenta are not soft. The arguments are
based on the fact that chiral logarithms come from soft parts
of the loop integrals and that these are well predicted as soon
as a sufficiently complete Lagrangian is used for the neighbourhood
of the relevant process and kinematics \citetwo{BC}{FS}.
I will discuss the application to $K\to\pi\pi$ \cite{BC}
and to a large set of formfactors:
the pion and kaon electromagnetic formfactor, the pion scalar formfactor,
$B,D\to\pi,H,\eta$ and $B\to D$ vector transitions \citetwo{BJ1}{BJ2}.
The arguments are checked using extra terms and/or different formalisms.
A two-loop check can be done for the pion electromagnetic and scalar formfactor.

A second area in which we made progress is the calculation
of leading logarithms to a much higher loop level than before in
a massive EFT. I will describe the principle behind the calculations and
results for the $O(N)$ sigma model \citetwo{BC1}{BC2}. We also get
the leading large $N$ results to all orders. Quantities discussed are the
mass, decay constant, vacuum expectation value, scalar and vector formfactors,
and meson-meson scattering.

The last part concerns the ongoing work on two-loop
calculations for symmetry breaking patterns other than standard ChPT.
Here we discuss the symmetry breaking patterns for the case of $N_F$ fermions
in a complex, real or pseudoreal gauge group representation.
Calculated so far are the mass, decay constant and vacuum expectation value
\cite{BL1} and meson-meson scattering \cite{BL2}. 
We founds some large
$N_F$ relations for scattering amplitudes between the three cases.
This work is relevant for models of dynamical symmetry breaking in
the standard model.

\setlength{\bibsep}{0.0em}
\begin{thebibliographynotitle}{99}

\bibitem{BC}
  J.~Bijnens, A.~Celis,
  Phys.\ Lett.\  {\bf B680 } (2009)  466-470.
  [arXiv:0906.0302]].

\bibitem{FS}
  J.~Flynn and C.~Sachrajda,
  Nucl. Phys.  {B812} (2009)  64.
  [arXiv:0809.1229].

\bibitem{BJ1}
  J.~Bijnens, I.~Jemos,
  Nucl.\ Phys.\  { B840 } (2010)  54-66.
  [arXiv:1006.1197].

\bibitem{BJ2}
  J.~Bijnens, I.~Jemos,
  Nucl.\ Phys.\  { B846 } (2011)  145-166.
  [arXiv:1011.6531].

\bibitem{BC1}
  J.~Bijnens, L.~Carloni,
  Nucl.\ Phys.\  { B827 } (2010)  237-255.
  [arXiv:0909.5086].

\bibitem{BC2}
  J.~Bijnens, L.~Carloni,
  Nucl.\ Phys.\  { B843 } (2011)  55-83.
  [arXiv:1008.3499].

\bibitem{BL1}
  J.~Bijnens, J.~Lu,
  JHEP { 0911 } (2009)  116.
  [arXiv:0910.5424].

\bibitem{BL2}
  J.~Bijnens, J.~Lu,
  JHEP { 1103 } (2011)  028.
  [arXiv:1102.0172].
\end{thebibliographynotitle}

\newabstract 
\begin{center}
{\large\bf Nucleon magnetic moments and electric polarizabilities with lattice QCD in background electric fields}\\[0.5cm]
Will Detmold$^{1,2}$, Brian Tiburzi$^{3}$ and {\bf Andr\'{e} Walker-Loud}$^4$,  
\\[0.3cm]
$^1$Dept. of Physics, College of William and Mary,
Williamsburg, VA 23187, USA\\
$^2$Jefferson Laboratory, Newport News, VA 23606, USA\\[0.3cm]
$^3$Center for Theoretical Physics, MIT, 
Cambridge, MA 02139, USA\\[0.3cm]
$^4$Lawrence Berkeley National Laboratory, 
Berkeley, CA 94720, USA\\[0.3cm]
\end{center}

The response of hadrons to low-energy electromagnetic probes is well described by the first few terms in a multi-pole expansion, which provide information about the hadrons internal structure.
Near the chiral limit, nucleon polarizabilities arise from the deformation of the charged pion cloud, and are highly constrained by effective interactions that emerge in the low-energy limit of QCD. Lattice QCD will play a crucial role in validating this low-energy picture, and the electromagnetic polarizabilities provide an area in which the lattice will influence phenomenology.

We describe a new method of computing charged and neutral hadron electromagnetic properties using lattice QCD and background fields.
Previously, we have shown how to determine the electric polarizabilities of charged and neutral pseudo scalar mesons with lattice QCD~\citetwo{Tiburzi:2008ma}{Detmold:2009dx}. 
We have recently extended this work to nucleons~\cite{Detmold:2010ts}.%
\footnote{With suitable choices of space and/or time varying background electromagnetic fields, one can compute the nucleon spin polarizabilites as well~\cite{Detmold:2006vu}.} 
Treating the spin relativistically is necessary to account for all effects at second order in the strength of the applied electric field. To determine the electric polarizability, we show that a background field analogue of the Born subtraction is necessary.
We demonstrate such a method, focussing on the determination of the nucleon magnetic moments.  Both the precision and values are consistent with independent methods utilizing momentum extrapolations of three-point functions~\cite{Hagler:2009ni}.

We conclude with a summary of current and future results as well as refinements which must be made to compare with experiment.

\setlength{\bibsep}{0.0em}
\begin{thebibliographynotitle}{99}
\bibitem{Tiburzi:2008ma} B.C.~Tiburzi, Nucl. Phys. A814 (2008) 74-108.
\bibitem{Detmold:2009dx} W.~Detmold, B.~C.~Tiburzi, A.~Walker-Loud, Phys.~Rev.~D79 (2009) 094505.
\bibitem{Detmold:2010ts} W.~Detmold, B.~C.~Tiburzi, A.~Walker-Loud, Phys.~Rev.~D81 (2010) 054502.
\bibitem{Detmold:2006vu} W.~Detmold, B.~C.~Tiburzi, A.~Walker-Loud, Phys.~Rev.~D73 (2006) 114505.
\bibitem{Hagler:2009ni} Ph.~H\"{a}gler, Phys. Rept. 490 (2010) 49-175.
\end{thebibliographynotitle}

\newabstract 
\begin{center}
{\large\bf Dispersive analysis of isospin breaking in $K_{\ell 4}$ decays}\\[0.5cm]
V\'eronique Bernard$^1$, {\bf S\'ebastien Descotes-Genon}$^{2}$, Marc Knecht$^3$\\[0.3cm]
$^1$ Institut de Physque Nucl\'eaire, F-91405 Orsay\\
$^2$  Laboratoire de Physique 
Th\'eorique, F-91405 Orsay\\
$^3$ Centre de Physique Th\'eorique,
CNRS-Luminy, Case 907, F-13288 Marseille\\[0.3cm]
\end{center}

The NA48/2 experiment has provided very accurate low-energy data on the difference of $S$- and $P$-wave $\pi\pi$ phase shifts through a careful analysis of $K^\pm\to \pi^+\pi^-e^\pm \nu$ decays~\cite{Batley:2010zza}. It was pointed out that this high level of accuracy required one to take into account isospin-breaking effects~\cite{Colangelo:2008sm}. The contributions from real and virtual photons can be treated (and have been removed) experimentally, estimating the Coulomb exchanges and incorporating radiative processes by a Monte-Carlo treatment. On the other hand, the effects from the mass differences between charged and neutral pions must be determined from a theoretical analysis.
The induced isospin-breaking corrections can be computed in the framework of chiral perturbation theory, assuming particular values of the $\pi\pi$ scattering lengths $a_0^0,a_0^2$.
But one can bypass these assumptions 
using an alternative and more general dispersive approach to reconstruct the amplitudes of interest~\cite{Knecht:1995tr}. 

The isospin-breaking corrections to Watson's theorem, responsible for the difference between the phases of the $K^\pm\to \pi^+\pi^-e^\pm \nu$ form factors  and those from $\pi\pi$ scattering in the isospin limit, fall into three categories: final-state phase-space difference, $\pi^0\pi^0\to \pi^+\pi^-$ rescattering, isospin-breaking in $K_{\ell 4}$ form factors between $\pi^+\pi^-$ and $\pi^0\pi^0$ final states. These isospin-breaking contributions induce also a dependence of the phase shifts on the leptonic invariant mass $s_l$.
Both $\pi\pi$ scattering amplitudes and $K_{\ell 4}$ form factors are reconstructed dispersively in order to compute the difference between the measured phases in $K^\pm\to \pi^+\pi^-e^\pm \nu$ and the corresponding phases in the isospin limit, at next-to-leading order, and at  first order in isospin breaking. 
We then assess these effects exploiting available estimates of  the relevant low-energy counterterms. We agree with existing works based on chiral perturbation theory~\cite{Colangelo:2008sm} when we take the $\pi\pi$ scattering lengths predicted in chiral perturbation theory. But when we vary $a_0^0,a_0^2$ in the universal band allowed by Roy equations, isospin-breaking corrections to the phase shifts
can change noticeably. These results will be used to reanalyse $K_{\ell 4}$ phase shifts to determine the pattern of
 $N_f=2$ chiral symmetry breaking~\cite{DescotesGenon:2001tn}.

\setlength{\bibsep}{0.0em}
\begin{thebibliographynotitle}{99}
\bibitem{Batley:2010zza}
  J.~R.~Batley {\it et al.}  [NA48-2 Collaboration],
  Eur.\ Phys.\ J.\  C {\bf 70} (2010) 635.
  
\bibitem{Colangelo:2008sm}
  G.~Colangelo, J.~Gasser and A.~Rusetsky,
  Eur.\ Phys.\ J.\  C {\bf 59}, 777 (2009).
  
\bibitem{Knecht:1995tr}
  M.~Knecht et al.,
  Nucl.\ Phys.\  B {\bf 457}, 513 (1995).

\bibitem{DescotesGenon:2001tn}
  S.~Descotes-Genon et al.  
  Eur.\ Phys.\ J.\  C {\bf 24} (2002) 469.
\end{thebibliographynotitle}

\newabstract 
\begin{center}
{\large\bf Quark mass dependence of meson-meson scattering: phases and
resonances from standard and unitarized Chiral Perturbation Theory }\\[0.5cm]
C. Hanhart$^1$, J. Nebreda$^2$, {\bf J.R. Pel\'aez}$^2$ and G. R\'{\i}os$^2$
 \\[0.3cm]
$^1$ Institut f\"ur Kernphysik (Theorie), Forschungzentrum J\"ulich, D-52425 J\"ulich, Germany \\[0.3cm]
$^2$ Dept. F\'{\i}sica Te\'orica II. Universidad Complutense, 28040, Madrid. Spain.
\end{center}

We use one and two-loop Chiral Perturbation Theory (ChPT) to study the
quark mass dependence of meson meson elastic scattering phase shifts 
and resonance poles. 
Phases are studied \cite{Nebreda:2011di} within SU(2) standard ChPT
and we show the soft mass dependence once phases are expressed in terms of momentum. 
We also show how the range of applicability can be extended by means of unitarization.
Resonances are generated through unitarized ChPT from a
dispersion relation in the form of the Inverse Amplitude Method,  
either within SU(2) to one loop \cite{Hanhart:2008mx} and to two loops \cite{Pelaez:2010fj},
or within the SU(3) formalism \cite{Nebreda:2010wv} to one loop. Details can also be found in the review \cite{Pelaez:2010er}. Here we show
a good agreement with existing lattice results on isospin 2 phases, the $\rho$ mass, $f_\pi$ and the highest isospin scattering lengths.
In addition we provide
predictions for $\rho$ couplings and all the parameters for the $f_0(600)$, $\kappa(800)$ 
and $K^*(892)$. 
The results may be used as a guideline for lattice studies and as insight on the
structure of the lightest hadronic resonances, particularly the
controversial light scalars.

\setlength{\bibsep}{0.0em}
\begin{thebibliographynotitle}{99}
\bibitem{Nebreda:2011di}
  J.~Nebreda, J.~R.~Pelaez, G.~Rios,
  [arXiv:1101.2171 [hep-ph]].

\bibitem{Hanhart:2008mx}
  C.~Hanhart, J.~R.~Pelaez, G.~Rios,
  Phys.\ Rev.\ Lett.\  {\bf 100}, 152001 (2008).
  [arXiv:0801.2871 [hep-ph]].

\bibitem{Pelaez:2010fj}
  J.~R.~Pelaez, G.~Rios,
  Phys.\ Rev.\  {\bf D82}, 114002 (2010).
  [arXiv:1010.6008 [hep-ph]].

\bibitem{Nebreda:2010wv}
  J.~Nebreda, J.~R.~Pelaez.,
  Phys.\ Rev.\  {\bf D81}, 054035 (2010).
  [arXiv:1001.5237 [hep-ph]].

\bibitem{Pelaez:2010er}
  J.~R.~Pelaez, J.~Nebreda, G.~Rios,
  Prog.\ Theor.\ Phys.\ Suppl.\  {\bf 186}, 113-123 (2010).
  [arXiv:1007.3461 [hep-ph]].
\end{thebibliographynotitle}

\newabstract 
\begin{center}
{\large\bf Coupled-channel approach to pion-nucleon-scattering}\\[0.5cm]
 {\bf Peter C. Bruns}$^1$, Maxim Mai$^2$,  Ulf-G. Mei{\ss}ner$^{2}$ \\[0.3cm]
$^1$Institut f\"ur Theoretische Physik,\\ Universit\"at Regensburg, D-93040 Regensburg, Germany\\[0.3cm]

$^2$HISKP (Theorie)
and Bethe Center for Theoretical Physics,\\
Universit\"at Bonn, D-53115 Bonn, Germany\\[0.3cm]

\end{center}

The analysis of meson-baryon scattering in the resonance region is a difficult non-perturbative problem for which, up to this date and to the best of our know\-ledge, only model treatments exist.
It is the aim of the work presented in this talk (see also \cite{bmm}) to see how far one can get in the description of meson-baryon scattering if one keeps a one-to-one correspondence of the terms in the scattering amplitude to dimensionally regularized Feynman graphs. This correspondence allows for a natural and straightforward method to implement gauge invariance in a chiral unitary framework for meson photoproduction, where the meson-baryon scattering amplitude discussed here serves as an ``extended vertex'' guaranteeing exact (two-particle) unitarity in the subspace of meson-baryon states. \\
To keep the abovementioned correspondence to Feynman graphs, we chose to apply the Bethe-Salpeter equation to iterate our potential, derived from the contact terms of the leading and next-to-leading order chiral meson-baryon Lagrangian, to infinite order.
 It is a clear drawback of the method under discussion that Born-terms can not be iterated, as the corresponding Feynman graphs would yield multiloop integrations which could only be treated numerically, introducing some cutoff regulator functions destroying the exact correspondence to Feynman graphs aimed at here.\\
As an example for the outcome of our model, we compare our results for the pion-nucleon s-wave scattering amplitudes to the data analysis provided by the SAID program. Though we fit only data up to 1.56 GeV, our description of the S11 partial wave above this energy works rather well. In particular, besides generating the $S_{11}$(1535) resonance, we also obtain one more resonance pole structure which can be associated with the $S_{11}$(1650). On the other hand, the $S_{31}$(1620) is not generated in our approach.\\

\setlength{\bibsep}{0.0em}
\begin{thebibliographynotitle}{99}
\bibitem{bmm}
P.~C.~Bruns, M.~Mai and U.-G.~Mei{\ss}ner,
  Phys.\ Lett.\  B {\bf 697} (2011) 254
  [arXiv:1012.2233 [nucl-th]].
\end{thebibliographynotitle}

\newabstract 
\begin{center}
\boldmath
{\large\bf Reconciling $J/\psi$ production at HERA, RHIC, Tevatron and LHC
with NRQCD factorization at next-to-leading order}\\[0.5cm]
Mathias Butensch\"on and {\bf Bernd A. Kniehl}\\[0.3cm]
II.\ Institut f\"ur Theoretische Physik, Universit\"at Hamburg,
Luruper Chaussee~149, 22761~Hamburg, Germany\\[0.3cm]
\end{center}
\unboldmath

We calculate the cross section of inclusive direct $J/\psi$ photoproduction
\cite{Butenschoen:2009zy} and hadroproduction \cite{Butenschoen:2010rq} at
next-to-leading order (NLO) within the factorization formalism of
nonrelativistic quantum chromodynamics (NRQCD) \cite{Bodwin:1994jh}, including
the full relativistic corrections due to the intermediate $^1\!S_0^{[8]}$,
$^3\!S_1^{[8]}$, and $^3\!P_J^{[8]}$ color-octet (CO) states.
We perform a combined fit of the CO long-distance matrix elements to the
transverse-momentum ($p_T$) distributions measured by CDF \cite{Acosta:2004yw}
at the Fermilab Tevatron and H1 \citetwo{Adloff:2002ex}{Aaron:2010gz} at DESY HERA
and demonstrate that they also successfully describe the $p_T$ distributions
from PHENIX \cite{Adare:2009js} at BNL RHIC and CMS \cite{Khachatryan:2010yr}
at the CERN LHC as well as the photon-proton c.m.\ energy and (with worse
agreement) the inelasticity distributions from H1
\citetwo{Adloff:2002ex}{Aaron:2010gz}.
This provides a first rigorous test of NRQCD factorization at NLO.
In all experiments, the CO processes are shown to be indispensable.

\setlength{\bibsep}{0.0em}
\begin{thebibliographynotitle}{99}
\bibitem{Butenschoen:2009zy}
  M.~Butensch\"on and B.~A.~Kniehl,
  Phys.\ Rev.\ Lett.\  {\bf 104} (2010) 072001
  [arXiv:0909.2798 [hep-ph]].
\bibitem{Butenschoen:2010rq}
  M.~Butensch\"on and B.~A.~Kniehl,
  Phys.\ Rev.\ Lett.\  {\bf 106} (2011) 022003
  [arXiv:1009.5662 [hep-ph]].
\bibitem{Bodwin:1994jh}
  G.~T.~Bodwin, E.~Braaten and G.~P.~Lepage,
  Phys.\ Rev.\  D {\bf 51} (1995) 1125
  [Erratum-ibid.\  D {\bf 55} (1997) 5853]
  [arXiv:hep-ph/9407339].
\bibitem{Acosta:2004yw}
  D.~Acosta {\it et al.}\  [CDF Collaboration],
  Phys.\ Rev.\  D {\bf 71} (2005) 032001
  [arXiv:hep-ex/0412071].
\bibitem{Adloff:2002ex}
  C.~Adloff {\it et al.}\  [H1 Collaboration],
  Eur.\ Phys.\ J.\  C {\bf 25} (2002) 25
  [arXiv:hep-ex/0205064].
\bibitem{Aaron:2010gz}
  F.~D.~Aaron {\it et al.}\  [H1 Collaboration],
  Eur.\ Phys.\ J.\  C {\bf 68} (2010) 401
  [arXiv:1002.0234 [hep-ex]].
\bibitem{Adare:2009js}
  A.~Adare {\it et al.}\  [PHENIX Collaboration],
  Phys.\ Rev.\  D {\bf 82} (2010) 012001
  [arXiv:0912.2082 [hep-ex]].
\bibitem{Khachatryan:2010yr}
  V.~Khachatryan {\it et al.}\  [CMS Collaboration],
  arXiv:1011.4193 [hep-ex].
\end{thebibliographynotitle}

\newabstract 
\begin{center}
{\large\bf What Top Mass is measured at the LHC ?}\\[0.5cm]
{\bf Andr\'e H. Hoang}$^{1,2}$, Iain W. Stewart$^3$ \\[0.3cm]
$^1$Fakult\"at f\"ur Physik, Universit\"at Wien, A-1090 Wien, Austria\vspace{0.5
  cm}\\[-5mm]
$^2$Max-Planck-Institut f\"ur Physik, F\"ohringer Ring 6, 80805 M\"unchen, Germany\vspace{0.5 cm}\\
$^3$Center for Theoretical Physics, Massachusetts Institute of Technology, Cambridge, MA 02139\\[3mm]

\end{center}

At LHC and the Tevatron the top mass is determined from comparing
measured distributions, such as the top invariant mass
distribution, with corresponding predictions from Monte-Carlo (MC) generators.  
Thus the quoted top mass is the MC mass parameter. In
principle, the MC top quark mass parameter corresponds to a specific mass scheme
(MC-scheme)
that depends on the way how the perturbative part of the event generator, i.e.\
the parton shower, is implemented and to which extend higher order matrix
element corrections are included.

Two arguments show that the top quark MC scheme is not
the pole scheme. First, the parton shower relies on a leading-log
approximation for soft and collinear radiation and, second, the shower
evolution 
is restricted from below by the shower
cutoff of $1$~GeV,
which acts as an IR cutoff for the perturbative
contributions. 
Virtual contributions needed to specify the pole scheme are absent and 
the IR cutoff protects the MC top mass scheme
from low-momentum renormalon contributions that plague the pole mass scheme. 
Up to now, real and virtual corrections needed to specify 
the MC scheme at ${\cal O}(\alpha_s)$ have not yet been quantified. This
restricts the application of the top mass in the MC scheme in precision physics.

Using Soft-Collinear-Effective-Theory it has recently become possible
to make first principle factorization predictions for jets initiated by heavy
quarks and define a jet mass scheme which the above issues. For $e^+e^-$
collisions the hemisphere jet mass 
and thrust distributions can be computed~\cite{Fleming:2007qr}. These results
allow the top mass to be determined from the 
experimental distributions with precise control of the scheme and of the
theoretical uncertainties. On the other hand, they could also be used to relate
the MC top quark mass scheme to the jet mass
scheme~\citetwo{Hoang:2008xm}{Hoang:2008yj}. To implement such a program for hadron 
collider MCs, corresponding factorization calculations with the jet mass still
have to be carried out.

\setlength{\bibsep}{0.0em}
\begin{thebibliographynotitle}{99}
\bibitem{Fleming:2007qr}
  S.~Fleming, A.~H.~Hoang, S.~Mantry and I.~W.~Stewart,
  Phys.\ Rev.\  D {\bf 77}, 074010 (2008),
  Phys.\ Rev.\  D {\bf 77}, 114003 (2008).
\bibitem{Hoang:2008xm}
  A.~H.~Hoang and I.~W.~Stewart,
  Nucl.\ Phys.\ Proc.\ Suppl.\  {\bf 185}, 220 (2008).
\bibitem{Hoang:2008yj}
  A.~H.~Hoang, A.~Jain, I.~Scimemi and I.~W.~Stewart,
  Phys.\ Rev.\ Lett.\  {\bf 101}, 151602 (2008).
\end{thebibliographynotitle}

\newabstract 
\begin{center}
{\large\bf Proper Heavy Quark Potential from Lattice QCD}\\[0.5cm]
{\bf Alexander Rothkopf}, Tetsuo Hatsuda, Shoichi Sasaki\\[0.3cm]
Department of Physics, The University of Tokyo,\\ 
Bunkyoku Hongo 7-3-1, 113-0031 Tokyo, Japan\\[0.3cm]
\end{center}

We refine our proposal \cite{Rothkopf:2009pk} on a non-perturbative derivation of the spin-independent complex potential for the two-body system consisting of a heavy quark and anti-quark at any temperature. This non-relativistic description is based on the separation of scales
\begin{equation}
 \frac{\Lambda_{\rm QCD}}{m_Qc^2}\ll1,\;\frac{T}{m_Qc^2}\ll1, \frac{{\bf p}}{m_Qc}\ll1, \label{Eq:scales}
\end{equation}
and utilizes the concept of quantum mechanical path integrals \cite{Barchielli:1986zs} for each heavy constituent. It yields the following expression for the static potential in terms of the spectral function of the thermal Wilson loop
\begin{equation}
 V^{(0)}(R,t) = -i\frac{\partial_t W_{\square}(R,t)}{W_{\square}(R,t)}=\frac{\int d\omega \, \omega\, e^{i\omega t} \rho_{\square}(\omega,R)}{\int d\omega\, e^{i\omega t} \rho_{\square}(\omega,R)}.\label{Eq:PropPot}
\end{equation}
If the spectral peak structure is well defined (e.g. Breit-Wigner or Gaussian type) a potential with real part (peak position) and
imaginary part (peak width) is obtained.

The analytical continuation of the real-time thermal Wilson loop to Euclidean times 
\begin{equation}
 W_{\square}(R,-it) = W_{\square}(R,\tau) = \int d\omega e^{-\omega \tau} \rho_{\square}(\omega,R) \label{Eq:SpecFunc2}
\end{equation}
allows us to infer the spectral functions required for the determination of the potential from lattice QCD simulations, using the Maximum Entropy Method\cite{Asakawa:2000tr}. A numerical evaluation in the quenched approximation $(T=0.78,1.17,2.33T_C)$ suggests that
the growth of the imaginary part, rather than the Debye screening of the real-part, will lead to a melting of bound states at temperatures above the deconfinement transition. 

\begin{thebibliographynotitle}{99}
\bibitem{Rothkopf:2009pk}
  A.~Rothkopf, T.~Hatsuda and S.~Sasaki,
  PoS {\bf LAT2009}, 162 (2009)
\bibitem{Barchielli:1986zs}
  A.~Barchielli, E.~Montaldi and G.~M.~Prosperi,
  Nucl.\ Phys.\  B {\bf 296}, 625 (1988)
\bibitem{Asakawa:2000tr}
  M.~Asakawa, T.~Hatsuda and Y.~Nakahara,
  Prog.\ Part.\ Nucl.\ Phys.\  {\bf 46}, 459 (2001)
\end{thebibliographynotitle}

\newabstract 
\begin{center}
{\large\bf The Polyakov Loop and Correlator of Polyakov Loops at NNLO}\\[0.5cm]
{\bf Antonio Vairo}\\[0.3cm]
Physik-Department, Technische Universit\"at M\"unchen,\\
James-Franck-Str. 1, 85748 Garching, Germany\\[0.3cm]
\end{center}

In \cite{Brambilla:2010xn}, we have studied 
the Polyakov loop and the correlator of two Polyakov loops at finite
temperature in the weak-coupling regime. 
We have calculated the Polyakov loop  in a colour representation $R$, $\langle L_R\rangle$, 
up to order $g^4$:
$$
\langle L_R\rangle=1+\frac{{C_R} \alpha_{\mathrm{s}} }{2}\frac{m_D}{T}+\frac{{C_R}\alpha_{\mathrm{s}}^2}{2}
\left[{C_A}\left(\ln\frac{m_D^2}{T^2}+\frac{1}{2}\right)-n_f\ln2\right]+\mathcal{O}(g^5).
$$
Our result disagrees with the old determination of \cite{Gava:1981qd}, for understood reasons, 
but agrees with the recent one of \cite{Burnier:2009bk}.
The calculation of the correlator of two Polyakov loops,  $C_{\mathrm{PL}}(r,T)$,
is performed at distances $r$ shorter than the inverse of the temperature $T$ 
and for electric screening masses $m_D$  larger than the Coulomb potential. 
In this regime, the calculation is new and accurate up to order $g^6(rT)^0$:
\begin{eqnarray*}
&& C_{\mathrm{PL}}(r,T) = 
\frac{N^2-1}{8N^2}\left\{ 
\frac{\alpha_{\mathrm{s}}(1/r)^2}{(rT)^2}
-2\frac{\alpha_{\mathrm{s}}^2}{rT}\frac{m_D}{T} \right.
+\frac{\alpha_{\mathrm{s}}^3}{(rT)^3}\frac{N^2-2}{6N}
\\
&&
\hspace{.5cm}
+ \frac{1}{2\pi }\frac{\alpha_{\mathrm{s}}^3}{(rT)^2}
\left(\frac{31}{9}{C_A}-\frac{10}{9}n_f +2\gamma_E\beta_0\right)
+ \frac{\alpha_{\mathrm{s}}^3}{rT}\Bigg[
{C_A}\left(-2 \ln\frac{m_D^2}{T^2} + 2-\frac{\pi^2}{4}\right) 
\\
&&
\hspace{.5cm}
\left.
+ 2n_f\ln 2\Bigg]
+\alpha_{\mathrm{s}}^2\frac{m_D^2}{T^2} -\frac{2}{9}\pi \alpha_{\mathrm{s}}^3 {C_A}
\right\}
+\mathcal{O}\left(g^6(rT),\frac{g^7}{(rT)^2}\right).
\end{eqnarray*}
In order to interpret the result, we have also evaluated the Polyakov-loop correlator in an effective 
field theory framework that takes advantage of the hierarchy of scales 
in the problem and makes explicit the non-relativistic bound-state dynamics \cite{Brambilla:2004jw}.
In the effective field theory framework, we show that the Polyakov-loop correlator is
at leading order in the multipole expansion the sum of a colour-singlet and a colour-octet 
quark-antiquark correlator, which are gauge invariant, 
and compute the corresponding colour-singlet and colour-octet free energies.

\setlength{\bibsep}{0.0em}
\begin{thebibliographynotitle}{99}

\bibitem{Brambilla:2010xn}
  N.~Brambilla, J.~Ghiglieri, P.~Petreczky and A.~Vairo,
  Phys.\ Rev.\  D {\bf 82} (2010) 074019
  [arXiv:1007.5172 [hep-ph]].

\bibitem{Gava:1981qd}
  E.~Gava and R.~Jengo,
  Phys.\ Lett.\  B {\bf 105} (1981) 285. 

\bibitem{Burnier:2009bk}
  Y.~Burnier, M.~Laine and M.~Veps\"al\"ainen,
  JHEP {\bf 1001} (2010) 054 
  [arXiv:0911.3480 [hep-ph]].

\bibitem{Brambilla:2004jw}
  N.~Brambilla, A.~Pineda, J.~Soto and A.~Vairo,
  Rev.\ Mod.\ Phys.\  {\bf 77} (2005)  1423 
  [arXiv:hep-ph/0410047].
\end{thebibliographynotitle}


\newabstract 
\begin{center}
{\large\bf How to identify hadronic molecules}\\[0.5cm]
{\bf Christoph Hanhart}\\[0.3cm]
Institut f\"{u}r Kernphysik, J\"ulich Center for Hadron
          Physics,  and
       Institute for Advanced Simulation, \\
          Forschungszentrum J\"{u}lich, D--52425 J\"{u}lich, Germany\\[0.3cm]
\end{center}
Many years ago Weinberg proposed a criterion that allows one to quantify from
data the molecular component of a bound state located in mass close to a 
continuum channel~\cite{SWein}. This criterion can, under certain conditions, be generalized to resonances~\cite{evidence}.
The central finding of this picture is that the effective coupling constant of
a state to the continuum channel becomes maximal, if the state is
predominantly a molecule made of the corresponding particles. 
In the talk I discuss the implications of this picture. E.g., within this
scheme it is possible to properly predict $f_0\to \gamma\gamma$~\cite{f02gamgam} under
the assumption that the $f_0(980)$ is a $\bar KK$ molecule. For another state,
the $D_s(2317)$ (as well as its spin partner $D^*_s(2460)$) one can predict
the hadronic width reliably, once it is assumed to be a $KD$ molecule~\cite{Dwidth}.
The same scheme also allows one to predict the quark mass dependence of
molecular states that could be checked using lattice QCD. Again for
$D_s(2317)$ and  $D^*_s(2460)$ we find a quite strong light quark mass or
equivalently pion mass
dependence of the mass of the states~\citetwo{phiDscat}{withmartin} -- since of a $c\bar s$ state the light quark mass dependence 
enters typically via loops and their contribution gets maximal when the
effective coupling gets maximal -- and an unusual, linear kaon mass
dependence of the masses. The latter is a natural consequence of the molecular
nature: the mass of the molecule
follows the threshold since the quark mass dependence of the binding
energy is weak~\cite{withmartin}.

\setlength{\bibsep}{0.0em}
\begin{thebibliographynotitle}{99}
\bibitem{SWein} S. Weinberg, Phys. Rev. {\bf
130}, 776 (1963); {\bf 131}, 440 (1963); {\bf 137} B672 (1965).
\bibitem{evidence}  V.~Baru et al.,
  Phys.\ Lett.\  {\bf B586}, 53-61 (2004).
\bibitem{f02gamgam}  C.~Hanhart, Y.~.S.~Kalashnikova, A.~E.~Kudryavtsev, A.~V.~Nefediev,
  Phys.\ Rev.\  {\bf D75 } (2007)  074015.
\bibitem{Dwidth}
  F.~-K.~Guo, C.~Hanhart, S.~Krewald, U.~-G.~Mei{\ss}ner,
  Phys.\ Lett.\  {\bf B666 } (2008)  251-255 and references therein.
\bibitem{phiDscat}
  F.~-K.~Guo, C.~Hanhart, U.~-G.~Mei{\ss}ner,
  Eur.\ Phys.\ J.\  {\bf A40 } (2009)  171-179.
\bibitem{withmartin}
M.~Cleven, F.~-K.~Guo, C.~Hanhart, U.~-G.~Mei{\ss}ner,
  Eur.\ Phys.\ J.\  {\bf A47 } (2011)  19.
\end{thebibliographynotitle}

\newabstract 
\begin{center}
{\large\bf Effective field theory approach to quarkonium at finite temperature}\\[0.5cm]
{\bf P\'eter Petreczky}\\[0.3cm]
Physics Deparment, Brookhaven National Laboratory,\\
Upton, NY, 11973 USA\\[0.3cm]
\end{center}

Quarkonium can be used to probe the properties of hot strongly interacting matter
produced in heavy ion collisions \cite{Rapp:2011zz}. Nevertheless first principle
calculations of quarkonium properties are still missing. Direct lattice calculations
of the quarkonium spectral functions turned out to be very difficult \cite{Datta:2004im}.
Attempts to study quarkonium properties using potential models have been made, but unlike 
in the zero temperature case the form of the potential is not known.
The effective field theory approach to quarkonium at finite temperature 
can address these problems \cite{Brambilla:2008cx}. As in the case of
zero temperature the quark anti-quark potential appears as a parameter
of the effective field theory Lagrangian. The form of the potential and
whether it is effected by the medium depends on the relation of the thermal
scales $T$ and $m_D$ (the Debye mass) and the zero temperature scales, the bound state
size $r$ and the binding energy $E_{bin}$. If $E_{bin} \gg T$ there are no
thermal correction to the potential. On the other hand if $E_{bin} \ll T$ there are thermal
corrections to the potential and their nature is different for $r \ll 1/T$ (short distance regime)
and for $r \gg 1/T$ (large distance regime). In the former case one first has to integrate out the scale
$1/r$ and then the scale $1/T$. Temperature effects show up as power law corrections in the potential
proportional to powers $T$ and $m_D$. In the later case one first integrates out the temperature scale,
which is equivalent to the hard thermal loop resummation and then the scale $1/r$. In the large distance
regime the potential is exponentially screened for $r \sim 1/m_D$ \cite{Brambilla:2008cx}. At
these distances the singlet $Q\bar Q$ potential equals to the singlet free energy calculated in 
\cite{Digal:2003jc}. Bottomonium properties at finite temperature 
have been calculated in the effective field theory approach assuming that 
calculations can be done in the short distance regime \cite{Brambilla:2010vq}.
For charmonium such an assumption is not
justified and the corresponding spectral functions can be calculated using  effective field theory
inspired potential model \cite{Miao:2010tk}.

\setlength{\bibsep}{0.0em}
\begin{thebibliographynotitle}{99}
\bibitem{Rapp:2011zz}
  R.~Rapp {\it et al.},
  Lect.\ Notes Phys.\  {\bf 814 } (2011)  335-529
\bibitem{Datta:2004im}
  S.~Datta {\it et al.},
  J.\ Phys.\ G {\bf G30 } (2004)  S1347-S1350.
\bibitem{Brambilla:2008cx}
  N.~Brambilla {\it et al.},
  Phys.\ Rev.\  {\bf D78 } (2008)  014017.

\bibitem{Digal:2003jc}
  S.~Digal, S.~Fortunato, P.~Petreczky,
  Phys.\ Rev.\  {\bf D68 } (2003)  034008.

\bibitem{Brambilla:2010vq}
  N.~Brambilla {\it et al.},
  JHEP {\bf 1009 } (2010)  038.

\bibitem{Miao:2010tk}
  C.~Miao, A.~M\'ocsy, P.~Petreczky,
 [arXiv:1012.4433 [hep-ph]].
\end{thebibliographynotitle}

\newabstract 
\begin{center}

\author{Y. Castin, C. Mora, L. Pricoupenko \\
\'{E}cole Normale Sup\'{e}rieure, CNRS and UPMC, 
Paris, France}

{\large\bf Four-body Efimov effect}\\[0.5cm]
{\bf Yvan Castin}$^1$, Christophe Mora$^2$, Ludovic Pricoupenko$^3$\\[0.3cm]
$^1$LKB and $^2$LPA, \'Ecole normale sup\'erieure, CNRS and UPMC, \\
24 rue Lhomond, 75231 Paris, France \\[0.3cm]
$^3$LPTMC, UPMC and CNRS, 4 place Jussieu, 75005 Paris, France \\[0.3cm]
\end{center}

The few-body problem with resonant two-body $s$-wave interaction
(infinite scattering length) can now be studied experimentally
with cold atoms. In particular, the three-body
Efimov phenomenon, consisting in the existence of an infinite number of trimer states with an asymptotically
geometric spectrum in the vicinity of a zero energy accumulation point \cite{Efimov},
has now experimental evidence with same spin bosons and with fermions in three different spin states \cite{manips_Efimov}.

On the contrary, the four-body Efimov effect has remained elusive. 
For same spin state bosons, as pointed out in \cite{Amado}, 
it is {\sl a priori} washed out by the
three-body Efimov effect:  Whereas low lying tetramers are possible \cite{Hammer4},
a tetramer state with an energy arbitrarily close to zero has eventually an energy larger than an Efimov trimer state
and decays into this trimer plus a free atom \cite{Deltuva}.

We have found a system where a four-body Efimov effect takes place \cite{nousPRL}: It is made of three same spin state
fermions of mass $M$ interacting only with a lighter particle of mass $m$.
The mass ratio
$\alpha=M/m$ can be used as a control knob: It was known that this system experiences
a three-body Efimov effect if and only if
$\alpha>\alpha_c(2;1)\simeq 13.607$ \citetwo{Efimov}{Petrov}. Using a combination of analytical arguments \cite{hidden}
and numerical solution of an integral equation, we show that an infinite number of Efimov tetramers exist over the interval of
mass ratio  $\alpha_c(3;1) < \alpha < \alpha_c(2;1)$, with
$\alpha_c(3;1)\simeq 13.384$. The four-body Efimov exponent $|s|$ is also calculated
as a function of $\alpha$ over that interval, and the experimental feasibility.
is discussed.

\setlength{\bibsep}{0.0em}
\begin{thebibliographynotitle}{99}

\bibitem{Efimov}
V. Efimov,
Sov. J. Nucl. Phys. {\bf 12}, 589 (1971);
Nucl. Phys. A {\bf 210}, 157 (1973);
A. Bulgac, V. Efimov, 
Sov. J. Nucl. Phys. {\bf 22}, 296 (1975).

\bibitem{manips_Efimov}
T. Kraemer {\sl et al.},
Nature {\bf 440} 315 (2006) 315;
T. Lompe {\sl et al.}, 
Science {\bf 330} (2010) 940.

\bibitem{Amado}
R. Amado, F. Greenwood, Phys. Rev. D {\bf 7} (1973) 2517.

\bibitem{Hammer4}
H.-W. Hammer, L. Platter, Eur. Phys. J. A {\bf 32} (2007) 113;
J. von Stecher, J.P. D'Incao, C.H. Greene, Nature Physics {\bf 5} (2009) 417.

\bibitem{Deltuva} 
A. Deltuva, Phys. Rev. A {\bf 82} (2010) 040701 (R).

\bibitem{nousPRL}
Y. Castin, C. Mora, L. Pricoupenko, Phys. Rev. Lett. {\bf 105} (2010) 223201.

\bibitem{Petrov}
D. Petrov, Phys. Rev. A {\bf 67} (2003) 010703.

\bibitem{hidden}
F. Werner, Y. Castin, Phys. Rev. A {\bf 74} (2006) 053604.
\end{thebibliographynotitle}

\newabstract 
\begin{center}
{\large\bf Universality in four-body scattering}\\[0.5cm]
{\bf A. Deltuva}\\[0.3cm]
Centro de F\'{\i}sica Nuclear da Universidade de Lisboa, P-1649-003 Lisboa, Portugal\\[0.3cm]
\end{center}

We study the four-particle system by solving exact 
Alt, Grassberger, and Sandhas (AGS) equations for the transition
operators \cite{grassberger:67} 
in the momentum-space framework
\cite{deltuva:07a}. The method  has  been applied successfully 
to the description of four-nucleon reactions
\citetwo{deltuva:07a}{deltuva:08a} and was extended to the four-boson system 
with large two-boson  scattering length $a$ \cite{deltuva:10c}.
In the latter case the three-boson  Efimov effect \cite{efimov:plb} 
has an impact on the the four-boson observables. 
The unitary limit ($a \to \infty$)
results for the atom-trimer scattering length, effective range, phase shifts, 
elastic and inelastic cross sections in reactions with highly 
excited trimers (at least 2nd excited state) 
were found to be related to the trimer binding energy in a universal way
\cite{deltuva:10c}.
Furthermore,  the existence of two tetramers 
for each Efimov trimer was predicted \citetwo{hammer:07a}{stecher:09a}.
However, only the two tetramers associated with the trimer
ground state are true bound states that have been studied in all the details
using standard bound state techniques \citetwo{hammer:07a}{stecher:09a}.
Higher tetramers are unstable bound states that lie above the lowest 
atom-trimer threshold
and for this reason their universal properties were far less known;
we determined their positions and widths  using proper scattering
calculations \cite{deltuva:11a}.
In contrast to \cite{stecher:09a}, we found that the shallow tetramer 
intersects the atom-trimer threshold twice and
in a particular regime becomes an inelastic virtual state. As a consequence,
 the atom-trimer scattering length exhibits  a resonant behaviour 
that might be observed as a resonant enhancement of the trimer relaxation
in the ultracold mixture of atoms and excited trimers.

\setlength{\bibsep}{0.0em}


\begin{thebibliographynotitle}{99}

\bibitem{grassberger:67}
P. Grassberger and W. Sandhas, Nucl. Phys. {\bf B2},  181  (1967); E. O. Alt,
  P. Grassberger, and W. Sandhas, JINR report No. E4-6688 (1972).

\bibitem{deltuva:07a}
A. Deltuva and A.~C. Fonseca, Phys.~Rev.~C {\bf 75},  014005  (2007);
Phys. Rev. Lett. {\bf 98},  162502  (2007).

\bibitem{deltuva:08a}
A. Deltuva, A.~C. Fonseca, and P.~U. Sauer, Phys.~Lett.~B {\bf 660},  471
  (2008).

\bibitem{deltuva:10c}
A. Deltuva, Phys.~Rev.~A {\bf 82},  040701(R)  (2010).

\bibitem{efimov:plb}
V. Efimov, Phys. Lett. B {\bf 33},  563  (1970).

\bibitem{hammer:07a}
H.-W. Hammer and L. Platter, Eur. Phys. J. A {\bf 32},  113  (2007).

\bibitem{stecher:09a}
J. von Stecher, J.~P. D'Incao, and C.~H. Greene, Nature Phys. {\bf 5},  417
  (2009).

\bibitem{deltuva:11a}
A. Deltuva, http://arxiv.org/abs/1103.2107.
\end{thebibliographynotitle}


\newabstract 
\begin{center}
{\large\bf Three bosons in two dimensions}\\[0.5cm]
{\bf Kerstin Helfrich} and Hans-Werner Hammer\\[0.3cm]
HISKP (Theorie)
and Bethe Center for Theoretical Physics,\\
Universit\"at Bonn, D-53115 Bonn, Germany\\[0.3cm]
\end{center}

In this talk I discussed two-dimensional bosonic gases exhibiting a large two-body scattering length. Within an effective field theory for resonant interactions, we calculate bound-state and scattering observables up to next-to-leading order, i.e.\ with the inclusion of the two-body effective range. We are especially interested in three-body observables such as the three-body binding energies, the atom-dimer scattering properties, and the three-body recombination rate. Significant effective range effects in the vicinity of the unitary limit are found and their implications are briefly discussed. More details can be found in Ref.~\cite{HH11}.

\setlength{\bibsep}{0.0em}
\begin{thebibliographynotitle}{99}
\bibitem{HH11} K.\ Helfrich and H.-W.\ Hammer, arXiv:1101.1891 [cond-mat.quant-gas].
\end{thebibliographynotitle}

\newabstract 
\begin{center}
{\large\bf Parametric Excitation of a 1D Gas in Integrable and Nonintegrable Cases}\\[0.5cm]
M. Colom\'e-Tatch\'e$^1$ and {\bf D.S.~Petrov$^{2,3}$}\\[0.3cm]
$^1$Groningen Bioinformatics Centre, Groningen Biomolecular Sciences and Biotechnology Institute,
University of Groningen, Nijenborgh 7, 9747 AG Groningen,
The Netherlands\\[0.3cm]
$^2$Laboratory LPTMS, CNRS, Universit\'{e} Paris Sud, 91405 Orsay, France\\[0.3cm]
$^3$RRC Kurchatov Institute, Kurchatov Square, 123182 Moscow, Russia\\[0.3cm]
\end{center}

Ultracold gases are ideal candidates for studies of fundamental differences between integrable and nonintegrable many-body dynamics. The main question is what measurement should one perform on a system in order to conclude on its integrability. The field of quantum chaos suggests to look at its spectral statistics \cite{Percival}. If energy levels are not correlated, we are dealing with an integrable or regular system. If, in contrast, levels repel each other, the system is not integrable. Another signature of integrability is the localization of eigenstates of a regular system in a certain physically meaningful basis, which suggests the dynamical probe of integrability: an excited initial state localized in this basis will stay localized during the temporal evolution.

In this work \cite{ColomePetrov} we compare responses of highly excited integrable and nonintegrable systems to an external time-dependent perturbation. We explore the idea that a perturbation localized in the same space as the eigenstates of the integrable system probes its local density of states, whereas in the nonintegrable case the states are delocalized, and the perturbation, no matter localized or not, couples all of them. Considering two 1D models on a ring we demonstrate that integrable systems can be much more stable with respect to slow variation of their Hamiltonian than nonintegrable ones. Namely, we consider the model of a single mobile impurity in a Fermi gas and the Lieb-Liniger model, and study their response to a periodic modulation of the coupling constant. This perturbation is localized in the many-body momentum space as it only changes the relative momentum of an atom pair. We show that the non-integrable system is sensitive to excitations with frequencies as low as the many-body mean level spacing, which is exponentially small, whereas the threshold frequency in the integrable case is much larger and scales polynomially with the system size.

\setlength{\bibsep}{0.0em}
\begin{thebibliographynotitle}{99}
\bibitem{Percival} I.~C. Percival, J. Phys. B: Atom. Molec. Phys. \textbf{6}, L229 (1973).
\bibitem{ColomePetrov} M. Colom\'e-Tatch\'e and D.S. Petrov, Phys. Rev. Lett. {\bf 106}, 125302 (2011).
\end{thebibliographynotitle}

\newabstract 
\begin{center}
{\large\bf Effective Field Theories in a Finite Volume}\\[0.5cm]
V\'eronique Bernard$^1$, Dominik Hoja$^2$, Michael Lage$^2$, Ulf-G. Mei{\ss}ner$^{2,3}$
and {\bf Akaki Rusetsky}$^2$\\[0.3cm]
$^1$Groupe de Physique Th\'eorique,
Universit\'e de Paris-Sud-XI/CNRS,\\ F-91406 Orsay, France\\[0.3cm]
$^2$HISKP (Theorie)
and Bethe Center for Theoretical Physics,\\
Universit\"at Bonn, D-53115 Bonn, Germany\\[0.3cm]
$^3$Forschungszentrum J\"ulich, J\"ulich Center for Hadron Physics,\\ Institut f\"ur Kernphysik 
and Institute for Advanced Simulation,\\ D-52425 J\"ulich, Germany
\\[0.3cm]
\end{center}

We review various applications of effective field theory methods in a
finite volume for the extraction of the characteristics of unstable particles 
from lattice QCD calculations. The talk is based on the material published
in the papers~\citetwo{scalar}{Hoja}, as well as on the results of  ongoing
investigations~\cite{Hoja-future}. 

As the first example, the scalar mesons
$f_0(980), a_0(980)$ are considered. We formulate criteria that 
distinguish, on one side, between hadronic molecules and tightly bound
quark states, and, on the other side, between  ordinary $q\bar q$ and  
tetraquark $qq\bar q\bar q $ states. Using these criteria will enable one to study
the nature of scalar mesons on the lattice~\cite{scalar}.

As the second example, we formulate a
procedure to calculate resonance matrix elements (the magnetic moments 
of resonances, the electromagnetic form factors, etc) on the lattice 
by using the background field method. In particular, we discuss
the problems which arise when the infinite-volume limit of these matrix elements
is performed. These problems have been analyzed in detail for the 
1+1-dimensional field theory by using a non-relativistic effective Lagrangian 
technique~\cite{Hoja}. Additional complications arise in the 3+1-dimensional 
case. In this talk,  ways to circumvent these complications have been 
discussed, see also Ref.~\cite{Hoja-future}.

\setlength{\bibsep}{0.0em}
\begin{thebibliographynotitle}{99}
\bibitem{scalar}
  V.~Bernard, M.~Lage, U.-G.~Mei{\ss}ner and A.~Rusetsky,
  JHEP 1101 (2011) 019.

\bibitem{Hoja}
  D.~Hoja, U.-G.~Mei{\ss}ner and A.~Rusetsky,
  JHEP 1004 (2010) 050.

\bibitem{Hoja-future}
D. Hoja {\it et al.,} in progress.
\end{thebibliographynotitle}

\newabstract 
\begin{center}
{\large\bf Extraction of light quark mass ratio from heavy quarkonia transitions}\\[0.5cm]
{\bf Feng-Kun Guo}\\[0.3cm]
Helmholtz-Institut f\"ur Strahlen- und Kernphysik (Theorie),\\
Universit\"at Bonn, D-53115 Bonn, Germany\\[0.3cm]
\end{center}

The $\psi'\to J/\psi\pi^0(\eta)$ was widely used in extracting the light quark
mass ratio $m_u/m_d$~\citethree{Ioffe:1979rv}{Donoghue:1992ac}{Leutwyler:1996qg}.
However, the extracted value using the PDG value for the measured branching
fractions is $0.39\pm0.02$. It is not compatible with the value obtained using
the pseudoscalar meson masses in chiral perturbation theory. We show that the
so-extracted quark mass ratio suffers from very large corrections from the
intermediate charmed meson loops~\cite{Guo:2009wr}. In fact, various
phenomenological calculations already suggested that heavy meson loops could
play an important role in the decays of heavy quarkonia (for an overview, see
\cite{Zhao:QNP}). Since $M_{c\bar c}-2 M_D\sim\Lambda_{\rm QCD}\ll M_D$, the
intermediate charmed mesons can be dealt with non-relativistically in a
non-relativistic effective field theory framework~\citetwo{Guo:2009wr}{Guo:2010ak}.
It was shown that the heavy meson-loop effects were also enhanced in the
transitions with the emission of one pion between two $P$-wave heavy quarkonia.
On the contrary, for the transitions between one $S$-wave and one $P$-wave heavy
quarkonia, the loops need to be analyzed case by case and often appear to be
suppressed~\cite{Guo:2010zk}. For a detailed power counting analysis, see
Ref.~\cite{Guo:2010ak}. In Ref.~\cite{Guo:2010ca}, we propose that the light
quark mass ratio can be extracted from the transitions $\Upsilon(4S)\to
h_b\pi^0(\eta)$, where the bottom meson loops are expected to be suppressed, to
be measured in the future.

\setlength{\bibsep}{0.0em}
\begin{thebibliographynotitle}{99}

\bibitem{Ioffe:1979rv}
  B.~L.~Ioffe,
  Yad.\ Fiz.\  29 (1979) 1611 [Sov. J. Nucl. Phys. 19 (1979) 827];
  B.~L.~Ioffe and M.~A.~Shifman,
  Phys.\ Lett.\  B 95 (1980) 99.

\bibitem{Donoghue:1992ac}
  J.~F.~Donoghue and D.~Wyler,
  Phys.\ Rev.\  D 45 (1992) 892.

\bibitem{Leutwyler:1996qg}
  H.~Leutwyler,
  Phys.\ Lett.\  B 378 (1996) 313
  [arXiv:hep-ph/9602366].

\bibitem{Guo:2009wr}
  F.~K.~Guo, C.~Hanhart and U.-G.~Mei{\ss}ner,
  Phys.\ Rev.\ Lett.\  103 (2009) 082003
  [Erratum-ibid.\  104 (2010) 109901]
  [arXiv:0907.0521 [hep-ph]].

\bibitem{Zhao:QNP}
   Y.~J.~Zhang, G.~Li and Q.~Zhao,
   Chin.\ Phys.\ C 34 (2010) 1181.

\bibitem{Guo:2010ak}
  F.~K.~Guo, C.~Hanhart, G.~Li, U.-G.~Mei{\ss}ner and Q.~Zhao,
  Phys.\ Rev.\  D 83 (2011) 034013
  [arXiv:1008.3632 [hep-ph]].

\bibitem{Guo:2010zk}
  F.~K.~Guo, C.~Hanhart, G.~Li, U.-G.~Mei{\ss}ner and Q.~Zhao,
  Phys.\ Rev.\  D 82 (2010) 034025
  [arXiv:1002.2712 [hep-ph]].

\bibitem{Guo:2010ca}
  F.~K.~Guo, C.~Hanhart and U.-G.~Mei{\ss}ner,
  Phys.\ Rev.\ Lett.\   105 (2010) 162001
  [arXiv:1007.4682 [hep-ph]].
\end{thebibliographynotitle}

\newabstract 
\begin{center}
{\large\bf Rescattering effects in \boldmath{$\eta$} and \boldmath{$\eta'$} decays}\\[0.5cm]
{\bf Bastian Kubis}\\[0.3cm]
HISKP (Theorie) and Bethe Center for Theoretical Physics,\\
Universit\"at Bonn, D-53115 Bonn, Germany\\[0.3cm]
\end{center}

Final-state interactions in meson decays into three particles  with small excess energy can be 
analyzed perturbatively in a variant of
non-relativistic effective field theory~\citetwo{CGKR}{NREFT}, originally developed
to study cusp effects
(see Ref.~\cite{FBproc} for a review of various applications).
Such rescattering effects are of high importance in $\eta\to3\pi$, which 
plays a central role in precision determinations of the light quark mass ratios.
We have analyzed the Dalitz plot parameters for this decay~\cite{SKDeta3pi} and in particular can 
reconcile the precise experimental values for the $\eta\to3\pi^0$ slope parameter $\alpha$ 
with seemingly unsuccessful chiral predictions~\cite{BGeta3pi}: 
we find $\alpha = -0.025\pm0.005$, to be compared with the current Particle Data Group average
$\alpha = -0.0317\pm0.0016$.  

A precise understanding of rescattering and the resulting \emph{imaginary parts}
of the decay amplitudes furthermore allows to compare $\alpha$  to 
the $\eta\to\pi^+\pi^-\pi^0$ Dalitz plot parameters in a consistent fashion~\cite{SKDeta3pi}:   
the latter's most recent experimental determination~\cite{KLOEcharged} 
seems to be inconsistent with the neutral decay channel.
Higher-order isospin violation~\cite{DKMeta3pi} plays only a minor role.

To obtain reliable descriptions of final-state interactions also at somewhat higher energies
and resum rescattering effects beyond perturbation theory, 
one has to resort to dispersion theoretical treatments \`a la Khuri and Treiman.
Applied to hadronic $\eta'$ decays,
these may, in addition to cusp effects~\cite{etaprime}, offer an alternative access to quark masses,
or allow to study $\pi\eta$ scattering.

\setlength{\bibsep}{0.0em}
\begin{thebibliographynotitle}{99}
\bibitem{CGKR}
  G.~Colangelo, J.~Gasser, B.~Kubis, A.~Rusetsky,
  Phys.\ Lett.\  B {\bf 638} (2006) 187
  [arXiv:hep-ph/0604084];

\bibitem{NREFT}
  J.~Gasser, B.~Kubis, A.~Rusetsky,
  arXiv:1103.4273 [hep-ph].

\bibitem{FBproc}
  B.~Kubis,
  EPJ Web Conf.\  {\bf 3} (2010) 01008
  [arXiv:0912.3440 [hep-ph]].

\bibitem{SKDeta3pi}
  S.~P.~Schneider, B.~Kubis, C.~Ditsche,
  JHEP {\bf 1102} (2011) 028
  [arXiv:1010.3946 [hep-ph]].

\bibitem{BGeta3pi}
  J.~Bijnens, K.~Ghorbani,
  JHEP {\bf 0711} (2007) 030
  [arXiv:0709.0230 [hep-ph]].

\bibitem{KLOEcharged}
  F.~Ambrosino {\it et al.}  [KLOE Collaboration],
  JHEP {\bf 0805} (2008) 006
  [arXiv:0801.2642 [hep-ex]].

\bibitem{DKMeta3pi}
  C.~Ditsche, B.~Kubis, U.-G.~Mei\ss ner,
  Eur.\ Phys.\ J.\  C {\bf 60} (2009) 83
  [arXiv:0812.0344 [hep-ph]].

\bibitem{etaprime}
  B.~Kubis, S.~P.~Schneider,
  Eur.\ Phys.\ J.\  C {\bf 62} (2009) 511
  [arXiv:0904.1320 [hep-ph]].
\end{thebibliographynotitle}

\newabstract 
\begin{center}
{\large\bf Low-Energy Pion-Photon Reactions and Chiral Symmetry}\\[0.3cm]
{\bf Norbert Kaiser}\\[0.2cm]
Physik-Department T39, Technische Universit\"at M\"unchen,\\
James Frank Str., 85747 Garching, Germany\\[0.3cm]
\end{center}
In this talk, I review the description of low-energy pion-Compton 
scattering in chiral perturbation theory. At one-loop order, the effects due
to the pion structure consist of the electric and magnetic polarizabilities 
(subject to the constraint $\alpha_\pi+\beta_\pi =0$) and a unique
pion-loop correction interpretable as photon scattering off the ``pion-cloud 
around the pion''. The latter compensates in the differential cross section 
$d\sigma/d\Omega_{\rm cm}$ partly the reduction effects due to the 
polarizabilities \cite{picross}. The two-loop corrections to charged 
pion-Compton scattering, first calculated by B\"urgi \cite{burgi} and recently
completed and reevaluated by Gasser et al.~\cite{gasser} are relatively small 
in the low-energy region $\sqrt{s}<4m_\pi$. Therefore chiral perturbation 
theory leads to firm predictions for the pion polarizabilities \cite{gasser}: 
$\alpha_\pi-\beta_\pi =(5.7\pm 1.0)\cdot 10^{-4}\,{\rm fm}^3$ and $\alpha_\pi+
\beta_\pi =0.16\cdot 10^{-4}\,{\rm fm}^3$. Because of the smallness of the 
pion-structure effects (less than $20\%$) radiative corrections of order 
$\alpha$ have to be included also in the analysis of pion-Compton scattering 
data. It is found that these QED radiative corrections \cite{radcor} have the 
same kinematical signature as the polarizability difference $\alpha_\pi-
\beta_\pi$, but their effects are suppressed in magnitude by about a factor 5 
or more. The other topic of my talk is the description of the (charged and 
neutral) pion-pair photoproduction processes $\pi^-\gamma \to \pi^- \pi^0\pi^0$
and $\pi^-\gamma \to\pi^+ \pi^-\pi^-$ at next-to-leading order in chiral 
perturbation theory \cite{dreipi}. Whereas the total cross section 
$\sigma_{\rm tot}(s)$ for the neutral channel $\pi^-\gamma \to \pi^- \pi^0\pi^0$ 
gets enhanced sizeably by the inclusion of chiral loop and counterterm 
corrections, the analogous effects turn out to be very small for the charged 
channel $\pi^-\gamma \to \pi^+ \pi^-\pi^-$. This different behavior can be 
understood from the varying influence of the chiral corrections on the 
pion-pion final state interaction ($\pi^+\pi^-\to \pi^0\pi^0$ versus $\pi^-\pi^-
\to \pi^-\pi^-$). The QED radiative corrections have also been calculated 
for the process $\pi^-\gamma \to \pi^- \pi^0\pi^0$ (simpler case) 
\cite{dreipi}. These affect the total cross section by at most $2\%$. 
The predictions of chiral perturbation theory for low-energy pion-photon 
reactions can be tested soon by the COMPASS experiment at CERN which uses 
Primakoff scattering of high-energy pions in the Coulomb-field of a 
heavy nucleus. A preliminary analysis for $\pi^-\gamma \to \pi^+ \pi^-\pi^-$ 
in the energy region $3m_\pi < \sqrt{s}<5m_\pi$ confirms the prediction of 
chiral perturbation theory. 
\vspace{-0.5cm}

\begin{thebibliographynotitle}{99}
\vspace{-0.3cm}
\bibitem{picross} N. Kaiser and J.M. Friedrich, Eur. Phys. J. A36 (2008) 181.\vs
\bibitem{burgi} U. B\"urgi,  Nucl. Phys. B479 (1996) 392; Phys. Lett. B377  
(1996) 147.\vs
\bibitem{gasser} J. Gasser, M.A. Ivanov and M.E. Sainio, Nucl. Phys. B745  
(2006) 84.\vs
\bibitem{radcor} N. Kaiser and J.M. Friedrich, Nucl. Phys. A812 (2008) 186.\vs
\bibitem{dreipi} N. Kaiser, Nucl. Phys. A848 (2010) 198; Eur. Phys. J. A46 
(2010) 373.\vs
\end{thebibliographynotitle}

\newabstract 
\begin{center}
{\large\bf The nuclear force problem: Have we finally reached the end of the tunnel?}
\\[0.5cm]
{\bf R. Machleidt}$^1$, E. Marji$^2$, and Ch. Zeoli$^1$,  
\\[0.3cm]
$^1$Department of Physics, University of Idaho, Moscow, Idaho 83844, USA
\\[0.3cm]
$^2$Physics Department, West Kentucky Community and Technical College,
Paducah,  Kentucky 42001, USA
\\[0.3cm]
\end{center}

In the past decade, there has been substantial progress in the derivation
of nuclear forces from chiral effective field theory (EFT). Accurate two-nucleon 
forces have been constructed at next-to-next-to-next-to-leading order (N$^3$LO) 
and applied [together with three-nucleon forces (3NFs) at NNLO] to nuclear 
few- and many-body 
systems---with a good deal of success. This may suggest that the 80-year old nuclear 
force problem has finally been cracked. Not so! Some pretty basic issues 
are still unresolved. In this talk, we focus on the two most pressing 
ones~\cite{ME10}, namely, sub-leading 3NFs and
the proper renormalization of the two-nucleon potential.

Concerning {\bf three-nucleon forces}, 
the bottom line is this:
\begin{itemize}
\item
The chiral 3NF at NNLO is insufficient.
\item
The chiral 3NF at N$^3$LO (in the $\Delta$-less theory) 
won't do it.
\item
However, sizable contributions are expected from one-loop 3NF diagrams
at N$^4$LO of the $\Delta$-less or N$^3$LO of the $\Delta$-full
theory. Thus, this is what needs to be attacked next.
\end{itemize}

{\bf Non-perturbative renormalization of the $NN$ potential.}
Naively, the `best' method of renormalization is the one where the momentum cutoff is taken to infinity while maintaining stable results. However, it has been shown for the chiral $NN$ potential that this type of renormalization leads to a rather erratic scheme of power counting and does not allow for a systematic order-by-order improvement of the predictions. This should not come as a surprise, since the chiral EFT these potentials are based upon is designed for momenta below the chiral-symmetry breaking scale of about 1 GeV. Therefore, in the spirit of an investigation which Lepage conducted in 1997 for a toy model, we have examined the cutoff dependence of the predictions by the chiral $NN$ potential at next-to-leading order (NLO) for phase shifts and $NN$ observables using cutoffs below the hard scale {\it identifying extended areas of cutoff independence} (``plateaus'').

\setlength{\bibsep}{0.0em}
\begin{thebibliographynotitle}{99}
\bibitem{ME10} 
R. Machleidt and D. R. Entem, J. Phys. G: Nucl. Part. Phys. {\bf 37} (2010) 064041.
\end{thebibliographynotitle}

\newabstract 
\begin{center}
{\large\bf Three-nucleon forces with explicit Delta fields}\\[0.5cm]
{\bf Hermann Krebs}\\[0.3cm]
Institut f\"ur Theor. Phys. II, Ruhr-Universit\"at Bochum,\\
44780 Bochum, Germany\\[0.3cm]
\end{center}

Description of light nuclei and low energy nuclear reactions can be given in
a model-independent way by using chiral effective field theory (EFT) of QCD. In this
framework nuclear forces are described by pion and nucleon (and possibly $\Delta$) rather
than fundamental quark-gluon degrees of freedom in harmony with the symmetries
of QCD. In the two-nucleon sector, chiral nuclear forces have been studied up to
next-to-next-to-next-to-leading order (N$^3$LO) in chiral expansion. All the
proton-neutron phase-shifts upto $E_{{\rm lab}}=200\,{\rm MeV}$ and deuteron
observables are very well described at this
order~\cite{Epelbaum:2008ga}. Three-nucleon forces have been analyzed up to
next-to-next-to-leading order (N$^2$LO) in chiral expansion. While many observables like e.g. differential cross sections 
of nucleon-deuteron scattering at low energies are very well described at
N$^2$LO there are some deficiencies in the description of vector analyzing
power in neutron-deuteron elastic scattering at low energies, known in the
literature as $A_y$-puzzle, and in the so called space-star configuration of
nucleon-deuteron break up (see~\cite{Epelbaum:2008ga} for extensive discussion). This suggests to
analyze the three-nucleon forces upto N$^3$LO.

Recently the expressions for long range part of N$^3$LO
contributions to the three-nucleon forces have been worked out by our
group~\cite{Bernard:2007sp} and partly by~\cite{Ishikawa:2007zz}. 
Their numerical implementations are still under development.

$\Delta(1232)$-resonance plays a special role in the nuclear physics. Its 
contributions to the nuclear forces are sizeable and are hidden in the
unnaturally large low energy constants in the $\Delta$-less chiral EFT. Taking
$\Delta$-resonance explicitly into account leads to a better convergence of the
chiral expansion of nuclear forces~\cite{Krebs:2007rh}. Our current analysis of the N$^3$LO long-range contributions 
to the three-nucleon force shows sizeable effects coming from $\Delta$-degrees of freedom. Sizable $\Delta$-contributions
to the three-nucleon force at N$^3$LO might potentially
 resolve the still existing puzzles in the three-nucleon
sector. It remains to be seen if the numerical implementations of these and other N$^3$LO-contributions
will lead to a better description of the experimental data.

\setlength{\bibsep}{0.0em}
\begin{thebibliographynotitle}{99}
\bibitem{Epelbaum:2008ga}
  E.~Epelbaum et al.
  Rev.\ Mod.\ Phys.\  {\bf 81}, 1773 (2009).
\bibitem{Bernard:2007sp}
  V.~Bernard et al.
  Phys.\ Rev.\  C {\bf 77}, 064004 (2008).
\bibitem{Ishikawa:2007zz}
  S.~Ishikawa and M.~R.~Robilotta,
  Phys.\ Rev.\  C {\bf 76}, 014006 (2007).
\bibitem{Krebs:2007rh}
  H.~Krebs et al.
  Eur.\ Phys.\ J.\  A {\bf 32}, 127 (2007).
\end{thebibliographynotitle}

\newabstract 
\begin{center}
{\large\bf What have we learned about three-nucleon forces at intermediate 
energies?}\\ [4mm]
{\bf N. Kalantar-Nayestanaki}\\ 
{\small \em KVI, University of Groningen, Groningen, The Netherlands}
\end{center}

Three-body systems have been studied in detail at KVI and other laboratories 
around the world in the last few years. 
Two categories of reactions have been chosen to investigate these systems,
namely elastic and break-up reactions in proton-deuteron scattering in which 
only hadrons are involved, and proton-deuteron capture reaction involving 
real and virtual photons in the final state. 
\setlength{\parindent}{8mm} 

Even though a relatively good understanding of most phenomena in nuclear 
physics has been arrived at by only considering two-nucleon forces, high 
precision three-nucleon data have revealed the shortcomings of these forces. 
Hadronic reactions in three-body systems excluding photons give a handle on 
effects such as those from three-body forces. In the last few decades, the two-nucleon 
system has been thoroughly investigated both experimentally and theoretically. 
These studies have resulted in modern potentials which describe the bulk of 
the data in a large range of energy. This knowledge can be employed in a 
Faddeev-like framework to calculate scattering observables in three-body systems. 
In regions and for the reactions in which the effects of Coulomb force are expected 
to be small or can be calculated accurately, and energies are low enough to avoid 
sizable relativistic effects, deviations from experimental data are a 
signature of three-body force effects. 


At KVI, various combinations of high-precision cross sections, analyzing 
powers and spin-transfer coefficients have been measured at different incident 
proton or deuteron beam energies between 100 and 200 MeV for a large range of 
scattering angles and for the reactions mentioned above. Calculations 
based on two-body forces only do not describe the data sufficiently. The inclusion of 
three-body forces improves the discrepancies with data significantly. However, there 
are still clear deficiencies in the calculations. 

The data from various laboratories have been combined to show globally how one 
can study the effects of three-body forces. It is very clear that the size of the effects
generally increases with 
increasing the beam energy. The predictions of cross sections generally improve 
when three-nucleon forces are explicitly added in the calculations. The spin 
observables, on the other hand, show a mixed picture. For some observables,
the predictions improve and for some other, they deviate even further from the 
data once the three-nucleon force is added. 

With the wealth of the data now available for various reactions in the three-body 
systems, nuclear forces should be developed which can describe the data. The 
developments within effective field theories should also be vigorously pursued to 
be able to predict observables at higher energies.


\newabstract 
\begin{center}
{\large\bf Nuclear physics from lattice effective field theory}\\[0.5cm]
Evgeny Epelbaum$^1$, Herman Krebs$^1$, {\bf Dean Lee}$^2$ 
\\and  Ulf-G. Mei{\ss}ner$^{3}$ \\[0.3cm]
$^1$Institut f\"ur Theor. Phys. II, Ruhr-Universit\"at Bochum,\\
44780 Bochum, Germany\\[0.3cm]
$^2$Department of Physics, North Carolina State University, \\
Raleigh, 27695, USA\\[0.3cm]
$^3$HISKP (Theorie)
and Bethe Center for Theor. Phys.,\\
Universit\"at Bonn, D-53115 Bonn, Germany\\[0.3cm]
\end{center}
Lattice effective field theory is a first principles method which combines the theoretical framework of effective field theory with supercomputer lattice simulations.  Some recent developments in lattice effective field theory can be found in Ref.~\citefour{Lee:2008fa}{Epelbaum:2009zs}{Epelbaum:2009pd}{Epelbaum:2010xt}.

Our collaboration has recently completed lattice calculations of the low-energy spectrum of carbon-12 using effective field theory \cite{Epelbaum:2011md}.  In addition to the ground state and excited spin-2 state, our calculations find a resonance at -85(3)~MeV with all of the properties of the Hoyle state and in agreement with the experimentally observed energy.  The Hoyle state plays a crucial role in the hydrogen burning of stars heavier than our sun and in the production of carbon and other elements necessary for life.  This excited state of the carbon-12 nucleus was postulated by Hoyle \cite{Hoyle:1954zz} as a necessary ingredient for the fusion of three alpha particles to produce carbon at stellar temperatures.  These lattice results provide insight into the structure of this unique state and new clues as to the amount of fine-tuning
needed in nature for the production of carbon in stars.

\setlength{\bibsep}{0.0em}
\begin{thebibliographynotitle}{99}
\bibitem{Lee:2008fa} D. Lee, Prog. Part. Nucl. Phys. 63 (2009) 117, arXiv:0804.3501 [nucl-th].
\bibitem{Epelbaum:2009zs} E. Epelbaum, H. Krebs, D. Lee, U.-G. Mei{\ss}ner, Eur. Phys. J. A41 (2009) 125, arXiv:0903.1666 [nucl-th].
\bibitem{Epelbaum:2009pd} E. Epelbaum, H. Krebs, D. Lee, U.-G. Mei{\ss}ner, Phys. Rev. Lett. 104 (2010) 142501, arXiv:0912.4195 [nucl-th].
\bibitem{Epelbaum:2010xt} E. Epelbaum, H. Krebs, D. Lee, U.-G. Mei{\ss}ner, Eur. Phys. J. A45 (2010) 335, arXiv:1003.5697 [nucl-th].
\bibitem{Epelbaum:2011md} E. Epelbaum, H. Krebs, D. Lee, U.-G. Mei{\ss}ner, arXiv:1101.2547 [nucl-th].
\bibitem{Hoyle:1954zz} F. Hoyle, Astrophys. J. Suppl. 1 (1954) 121.
\end{thebibliographynotitle}

\newabstract 
\begin{center}
{\large\bf The hadronic vacuum polarisation contribution to $(g-2)_\mu$
from lattice QCD}\\[0.5cm]
Michele Della Morte$^{1}$, Benjamin J\"ager$^1$, Andreas J\"uttner$^2$ 
\\and  {\bf Hartmut Wittig}$^{1}$ \\[0.3cm]
$^1$Institut f\"ur Kernphysik and
Helmholtz-Institut Mainz, University of Mainz, D-55099 Mainz, Germany\\[0.3cm]
$^2$CERN, Theory Division, CH-1211 Geneva 23, Switzerland\\[0.3cm]
\end{center}

The anomalous magnetic moment of the muon, $a_\mu={1\over2}(g-2)_\mu$,
provides one of the most stringent tests for physics beyond the
Standard Model, thanks to the impressive precision which has been
reached in its direct experimental measurement as well as in
theoretical predictions \cite{Jegerlehner:2009ry}. The accuracy of the
latter is limited by hadronic contributions, notably hadronic vacuum
polarisation. We report on our efforts to evaluate this contribution,
$a_\mu^{\rm{had}}$, using lattice simulations of QCD. The starting
point is the convolution integral
\begin{equation}
  a_\mu^{\rm had} = 4\pi^2\left(\frac{\alpha}{\pi}\right)^2
  \int_0^{\infty} {d}Q^2\,f(Q^2)\big\{ \Pi(Q^2)-\Pi(0) \big\},
\label{eq:amuhad}
\end{equation}
where $\Pi(Q^2)$ is related to the vacuum polarisation tensor. Since
$f(Q^2)$ is strongly peaked for momenta near $m_\mu$, which is an
order of magnitude smaller than what can conventionally be realised,
lattice simulations cannot constrain the integral where it receives
its dominant contribution. Furthermore, the Wick contractions for the
vacuum polarisation tensor produce contributions from so-called
quark-disconnected diagrams, which are notoriously difficult to
compute with good statistical accuracy. Our calculation involves
several novel ideas: using partially quenched ChPT, it was shown in
\cite{MorteJuett10} that connected and disconnected contributions can
be evaluated separately. Furthermore, the latter can be shown to be
suppressed by a factor~10. During the first stage of the project we
therefore concentrate on the connected contribution, which is
evaluated using twisted boundary conditions \cite{twbc}. This shifts
the minimum accessible value of $Q$ into the peak region of the
convolution function. Our preliminary results
\citetwo{wittig_lat10}{juettner_conf9} indicate that our strategy leads to
a significant improvement of the overall accuracy in lattice
calculations of $a_\mu^{\rm had}$.

\vspace*{-0.3cm} 
\setlength{\bibsep}{0.0em}
\begin{thebibliographynotitle}{99}
\vspace*{-0.3cm} 
\bibitem{Jegerlehner:2009ry}
  F.~Jegerlehner, A.~Nyffeler,
  Phys. Rept. 477 (2009) 1.
\bibitem{MorteJuett10}
  M.~Della Morte and A.~J\"uttner,
  JHEP 1011 (2010) 154.
\bibitem{twbc}
  J.M. Flynn, A.~J{\"u}ttner and C.T. Sachrajda,
  Phys. Lett. B632 (2006) 313.
\bibitem{wittig_lat10}
  B.B.~Brandt et al.,
  PoS LATTICE2010 (2010) 164, arXiv:1010.2390.
\bibitem{juettner_conf9}
  M.~Della Morte, B.~J\"ager, A.~J\"uttner and H.~Wittig,
  arXiv:1011.5793.
\end{thebibliographynotitle}


\newabstract 
\begin{center}
{\large\bf Prediction of super-heavy $N^*$ and $\Lambda^*$ resonances \\ with hidden charm and beauty}\\[0.5cm]
Jia-Jun Wu$^{1,3}$, {\bf Bing-Song Zou}$^{1,3}$, Raquel
Molina$^{2,3}$ and  Eulogio Oset$^{2,3}$ \\[0.3cm]
$^1$ Institute of High Energy Physics, CAS, Beijing 100049, China\\[0.3cm]
$^2$ Dept. de Fisica Teorica and IFIC, Univ. de Valencia, 46071 Valencia, Spain\\[0.3cm]
$^3$ Theoretical Physics Center for Science Facilities, CAS, Beijing 100049, China\\[0.3cm]
\end{center}

The interaction between various charmed mesons and charmed baryons
are studied within the framework of the coupled channel unitary
approach with the local hidden gauge formalism~\cite{Wujj1}. Several
meson-baryon dynamically generated narrow $N^*$ and $\Lambda^*$
resonances with hidden charm are predicted with mass around 4.3 GeV
and width smaller than 100 MeV. The predicted new resonances
definitely cannot be accommodated by quark models with three
constituent quarks and can be looked for at the forthcoming
PANDA/FAIR experiments.

Then the same approach is extended to the hidden beauty
sector~\cite{Wujj2}. A few narrow $N^*$ and $\Lambda^*$ around 11
GeV are predicted as dynamically generated states from the
interactions of heavy beauty mesons and baryons. Production cross
sections of these predicted resonances in $pp$ and $ep$ collisions
are estimated as a guide for the possible experimental search at
relevant facilities.

The S-wave $\Sigma_c\bar D$ and $\Lambda_c\bar D$ states with
isospin I=1/2 and spin S=1/2 are also dynamically investigated
within the framework of a chiral constituent quark model by solving
a resonating group method (RGM) equation by W.L.Wang et
al.~\cite{Wangwl}. They confirm that the interaction between
$\Sigma_c$ and $\bar D$ is attractive and results in a $\Sigma_c\bar
D$ bound state not far below threshold.

\setlength{\bibsep}{0.0em}
\begin{thebibliographynotitle}{99}
\bibitem{Wujj1}
  J.~J.~Wu, R.~Molina, E.~Oset and B.~S.~Zou,
  Phys.\ Rev.\ Lett.\  {\bf 105} (2010) 232001;
  arXiv:1011.2399 [nucl-th].
\bibitem{Wujj2}
  J.~J.~Wu and B.~S.~Zou,
  arXiv:1011.5743 [hep-ph].
\bibitem{Wangwl}
  W.~L.~Wang, F.~Huang, Z.~Y.~Zhang and B.~S.~Zou,
  arXiv:1101.0453 [nucl-th].
\end{thebibliographynotitle}

\newabstract 
\begin{center}
{\large\bf A coherent view of the charmonium hadronic and radiative decays}\\[0.5cm]
{\bf Qiang Zhao}$^{1,2}$
\\[0.3cm]
$^1$ Institute of High Energy Physics, Chinese Academy of
Sciences, Beijing 100049, P.R. China \\
$^2$ Theoretical Physics Center for Science Facilities, CAS, Beijing
100049, China
\\[0.3cm]
\end{center}

The open charm effects via intermediate hadron loop transitions seem
to be essential for understanding some of those long-standing
puzzles in charmonium decays into light hadrons. Some of those
include the $\psi(3770)$ non-$D\bar{D}$ decays, and the so-called
``$\rho\pi$ puzzle'' which is about the abnormally small ratio of
$Br(\psi^\prime\to\rho\pi)/Br(J/\psi\to \rho\pi)$ in comparison with
the pQCD expected ``12\%", namely, the ``12\% rule''. We show by
quantitative calculations that the intermediate charmed meson loops
provide a correlative mechanism for the Okubo-Zweig-Iizuka (OZI)
rule violations and helicity selection rule violations in charmonium
decays into light
hadrons~\citefour{Zhang:2009kr}{Li:2007ky}{Liu:2009vv}{Liu:2010um}. Further
possible evidences for the intermediate charmed meson loops in the
isospin violating transitions between charmonium
states~\citetwo{Guo:2010zk}{Guo:2010ak}, and radiative
decays~\citetwo{Li:2007xr}{Zhao:2010mm} are also discussed.

\setlength{\bibsep}{0.0em}
\begin{thebibliographynotitle}{99}
\bibitem{Zhang:2009kr}
  Y.~J.~Zhang, G.~Li and Q.~Zhao,
  Phys.\ Rev.\ Lett.\  {\bf 102}, 172001 (2009)
  [arXiv:0902.1300 [hep-ph]].

\bibitem{Li:2007ky}
  G.~Li, Q.~Zhao and C.~H.~Chang,
  J.\ Phys.\ G {\bf 35}, 055002 (2008)
  [arXiv:hep-ph/0701020].

\bibitem{Liu:2009vv}
  X.~H.~Liu and Q.~Zhao,
  Phys.\ Rev.\  D {\bf 81}, 014017 (2010)
  [arXiv:0912.1508 [hep-ph]].

\bibitem{Liu:2010um}
  X.~H.~Liu and Q.~Zhao,
  J.\ Phys.\ G {\bf 38}, 035007 (2011)
  [arXiv:1004.0496 [hep-ph]].

\bibitem{Guo:2010zk}
  F.~K.~Guo, C.~Hanhart, G.~Li, U.-G.~Mei{\ss}ner and Q.~Zhao,
  Phys.\ Rev.\  D {\bf 82}, 034025 (2010)
  [arXiv:1002.2712 [hep-ph]].

\bibitem{Guo:2010ak}
  F.~K.~Guo, C.~Hanhart, G.~Li, U.-G.~Mei{\ss}ner and Q.~Zhao,
  Phys.\ Rev.\  D {\bf 83}, 034013 (2011)
  [arXiv:1008.3632 [hep-ph]].

\bibitem{Li:2007xr}
  G.~Li and Q.~Zhao,
  Phys.\ Lett.\  B {\bf 670}, 55 (2008)
  [arXiv:0709.4639 [hep-ph]].

\bibitem{Zhao:2010mm}
  Q.~Zhao,
  Phys.\ Lett.\  B {\bf 697}, 52 (2011)
  [arXiv:1012.1165 [hep-ph]].
\end{thebibliographynotitle}

\newabstract 
\begin{center}
{\large\bf Dressed Polyakov loop and the relation between chiral and deconfinement phase transitions}\\[0.5cm]
Fukun Xu$^1$, Tamal K. Mukherjee$^{1,2}$, {\bf Mei Huang}$^{1,2}$ \\[0.3cm]
$^{1}$ Institute of High Energy Physics, Chinese Academy of
Sciences, Beijing, China \\[0.3cm]
$^{2}$ Theoretical Physics Center for Science
Facilities, Chinese Academy of Sciences, Beijing, China\\[0.3cm]

\end{center}

The interplay between chiral restoration and deconfinement phase
transition is very important for the QCD phase diagram at high
temperature and density. Recently the proposed ``quarkayonic" phase
\cite{quarkayonic} raises much interests in discussing the
possibility of chiral symmetric but confined phase structure. In the
Polyakov-loop NJL model \cite{PNJL}, the relation between the two
phase transitions is much parameter dependent, the chiral
restoration can happen earlier or latter than the deconfinement
phase transition, also the two phase transitions can coincide with
each other. We use the dressed Polyakov loop \cite{dpl} as an
equivalent order parameter of deconfinement phase transition, and
investigate the chiral and deconfinement phase transitions. In Ref.
\cite{our} we find  that in the case of first order and second order
phase transitions, the chiral phase transition always coincide with
deconfinement phase transition, and in the case of crossover, the
chiral transition temperature is always smaller than that of the
deconfinement. We also find that the phase transitions for light
$u,d$ quarks and $s$ quark are sequentially happened. Our result at
zero baryon density agrees with the lattice result from
Wuppetal-Budapest group \cite{lattice}.

\setlength{\bibsep}{0.0em}
\begin{thebibliographynotitle}{99}
\bibitem{quarkayonic}
 L.~McLerran and R.~D.~Pisarski,
  Nucl.\ Phys.\  A {\bf 796}, 83 (2007).
\bibitem{PNJL}
  C.~Ratti, M.~A.~Thaler and W.~Weise,
  Phys.\ Rev.\  D {\bf 73}, 014019 (2006);
 C.~Sasaki, B.~Friman and K.~Redlich,
  Phys.\ Rev.\  D {\bf 75}, 074013 (2007);
W.~j.~Fu, Z.~Zhang and Y.~x.~Liu,
  Phys.\ Rev.\  D {\bf 77}, 014006 (2008);
 K.~Fu\-ku\-shi\-ma,
  Phys.\ Rev.\  D {\bf 77}, 114028 (2008),
  [Erratum-ibid.\  D {\bf 78}, 039902 (2008)].
\bibitem{dpl}
  C.~Gattringer,
  Phys.\ Rev.\ Lett.\  {\bf 97}, 032003 (2006).
\bibitem{our} T.~K.~Mukherjee, H.~Chen and M.~Huang,
  Phys.\ Rev.\  D {\bf 82}, 034015 (2010);
   F.~Xu, T.~K.~Mukherjee, H.~Chen and M.~Huang,
  arXiv:1101.2952 [hep-ph].
\bibitem{lattice} Y.~Aoki, Z.~Fodor, S.~D.~Katz and K.~K.~Szabo,
  Phys.\ Lett.\  B {\bf 643}, 46 (2006).
\end{thebibliographynotitle}

\newabstract 
\begin{center}
{\large\bf Heavy Quarkonium in a weakly coupled Quark-Gluon Plasma}\\[0.5cm]
{\bf Jacopo Ghiglieri}$^{1,2}$\\[0.3cm]
$^1$Phys.-Dep. T30f, Technische Universit\"at M\"unchen,\\
James-Franck-Str. 1, 85748 Garching, Germany\\[0.3cm]
$^2$Excellence Cluster Universe, Technische Universit\"at M\"unchen,\\ 
Boltzmannstr. 2, 85748, Garching, Germany\\[0.3cm]
\end{center}
Heavy quarkonium has been suggested since long ago \cite{Matsui:1986dk} as a probe of the medium that forms in heavy-ion collision
experiments.
Recent efforts have focused on a systematic derivation of the heavy quark potential at finite temperature. To this end 
an Effective Field Theory (EFT) framework was constructed, based on the successful non-relativistic EFTs at $T=0$ for heavy quarkonium \cite{Brambilla:2004jw}.
This non-relativistic EFT study of quarkonium bound states in a plasma was initiated in  \cite{Brambilla:2009cd} in the static limit, and in \cite{Escobedo:2008sy} for QED.\\
In this talk we report the main findings of \cite{Brambilla:2010vq}, where this systematic and rigorous EFT approach is employed  for the calculation of the heavy quarkonium spectrum and decay
  width in a QCD plasma, whose temperature $T$ and screening
  mass $m_D\sim gT$ satisfy the hierarchy $m\alpha_\mathrm{s} \gg T \gg m \alpha_\mathrm{s}^2 \gg m_D$
  ($m$ being the heavy-quark mass, $m\alpha_\mathrm{s}$ the typical momentum transfer and $m\alpha_\mathrm{s}^2$ the binding energy). We first
  sequentially integrate out the scales $m$, $m\alpha_\mathrm{s}$ and $T$ and next we carry out the calculations in the resulting effective  theory. We finally discuss the relevance of this hierarchy and the implications of our results concerning the behaviour of the ground states of bottomonium in heavy ion collisions.

\setlength{\bibsep}{0.0em}
\begin{thebibliographynotitle}{99}
	\bibitem{Matsui:1986dk}
	  T.~Matsui and H.~Satz,
	  Phys.\ Lett.\  B {\bf 178}, 416 (1986).
	\bibitem{Brambilla:2004jw}
	  N.~Brambilla, A.~Pineda, J.~Soto and A.~Vairo,
	  Rev.\ Mod.\ Phys.\  {\bf 77}, 1423 (2005) 
	  [arXiv:hep-ph/0410047].
	\bibitem{Brambilla:2009cd}
	  N.~Brambilla, J.~Ghiglieri, A.~Vairo and P.~Petreczky,
	  Phys.\ Rev.\  D {\bf 78}, 014017 (2008) 
	  [arXiv:0804.0993 [hep-ph]].
		\bibitem{Escobedo:2008sy}
		  M.~A.~Escobedo and J.~Soto,
		  Phys. Rev. A {\bf 78}, 032520 (2008), [arXiv:0804.0691 [hep-ph]],
		  Phys.\ Rev.\ A {\bf 82}, 042506 (2010),	  [arXiv:1008.0254 [hep-ph]].
	\bibitem{Brambilla:2010vq}
	  N.~Brambilla, M.~A.~Escobedo, J.~Ghiglieri, J.~Soto, A.~Vairo,
	  JHEP {\bf 1009 } (2010)  038.
	  [arXiv:1007.4156 [hep-ph]].
\end{thebibliographynotitle}


\newabstract 
\begin{center}
{\large\bf Some novel developments in quarkonium electromagnetic transitions}\\[0.5cm]
{\bf Yu Jia}\\[0.3cm]
Institute of High Energy Physics and TPCSF,\\  
Chinese Academy of Sciences,
Beijing 100049, China\\[0.3cm]
\end{center}

Some recent theoretical developments in understanding quarkonium radiative
transitions have been presented in this talk. 1). The recent BaBar experiment
indicated that the $X(3872)$ charmonium state may carry the quantum number $2^{-+}$,
instead of the commonly assumed $1^{++}$\cite{delAmoSanchez:2010jr}. 
By assuming the $X$ meson to be the canonical $\eta_{c2}$ meson, 
we comprehensively investigate the spin-flipped radiative transitions 
$\eta_{c2}\to J/\psi(\psi')+\gamma$\cite{Jia:2010jn}
by utilizing potential nonrelativistic QCD (pNRQCD) framework\cite{Brambilla:2005zw} 
and phenomenological potential models. 
We have considered the $^3S_1-^3D_1$ mixing effects in $\psi'$ and
identified all the three multipole amplitudes. Comparing our predictions with the existing 
$B$ factory measurements, we tend to conclude that the assignment of the $X(3872)$ with the
$\eta_{c2}$ seems to be strongly disfavored.

2). The long-awaited bottomonium ground state $\eta_b(1S)$ has been recently discovered in hindered magnetic
dipole transitions process $\Upsilon(3S)\to \eta_b+\gamma$\cite{Aubert:2008vj}. In such a case, the emitted 
photon carries a rather large energy about $1$ GeV, which casts some doubt on the applicability of the conventional 
long-wavelength approximation.
By assigning the photon momentum as semi-hard ($k\sim {\cal O}(mv)$),  where 
$mv$ signifies the typical heavy quark 3-momentum inside quarkonium, we have developed a novel {\it hard-scattering} mechanism, 
to describe these kinds of strongly hindered radiative transitions\cite{Jia:2009yg}. 
The reasonable agreement with the BaBar measurement has been achieved.

\setlength{\bibsep}{0.0em}
\begin{thebibliographynotitle}{99}
\bibitem{delAmoSanchez:2010jr}
  P.~del Amo Sanchez {\it et al.}  [BABAR Collaboration],
  Phys.\ Rev.\  D {\bf 82}, 011101 (2010).
\bibitem{Jia:2010jn}
  Y.~Jia, W.~L.~Sang and J.~Xu, arXiv:1007.4541 [hep-ph].
\bibitem{Brambilla:2005zw}
  N.~Brambilla, Y.~Jia and A.~Vairo,
 Phys.\ Rev.\  D {\bf 73}, 054005 (2006).
\bibitem{Aubert:2008vj}
  B.~Aubert {\it et al.}  [BABAR Collaboration], Phys.\ Rev.\ Lett.\  {\bf 101}, 071801 (2008)
  [Erratum-ibid.\  {\bf 102}, 029901 (2009)].
\bibitem{Jia:2009yg} Y.~Jia, J.~Xu and J.~Zhang, Phys.\ Rev.\  D {\bf 82}, 014008 (2010).
\end{thebibliographynotitle}

\newabstract 
\begin{center}
{\large\bf Electric properties of halo nuclei from EFT}\\[0.5cm]
{\bf Daniel Phillips}$^{1,2}$ and Hans-Werner Hammer$^2$ 
 \\[0.3cm]
$^1$Department of Physics and Astronomy, Ohio University,
Athens, OH 45701, USA\\[0.2cm]
$^2$HISKP (Theorie)
and Bethe Center for Theoretical Physics,
Universit\"at Bonn, D-53115 Bonn, Germany\\[0.2cm]

\noindent
{\small Supported by the US DOE, the BMBF, and the Mercator programme of the DFG.}
\end{center}

\vskip -0.1cm

Our recent computation of the electromagnetic properties of single-neutron halo nuclei \cite{HP} uses ``halo effective field theory"~\cite{BHvK}. The expansion in halo EFT is in powers of $R_{\rm core}/R_{\rm halo}$, where $R_{\rm core}$ and $R_{\rm halo}$ are the range of the core-neutron interaction and the size of the bound state, respectively.
Carbon-19 is one nucleus to which we apply our theory: it has a 1/2$^+$ state which is predominantly a ${}^{18}$C core with an (s-wave) orbiting neutron. In that case the Coulomb dissociation data of Ref.~\cite{Nak1} are quite well described by an NLO computation in halo EFT. The $d{\rm B(E1)}/dE$ strength measured in such experiments has universal features. 

The Beryllium-11 nucleus has more low-energy levels: there a shallow 1/2$^+$ (s-wave) and 1/2$^-$ (p-wave) state are present. At LO in halo EFT we have three parameters: the binding energies of these two states, as well as the effective ``range" for p-wave ${}^{10}$Be-n scattering. We use data on the two levels and the B(E1)
strength of the transition between them to fix these parameters.

The $d{\rm B(E1)}/dE$ spectrum obtained from Coulomb excitation of ${}^{11}$Be into ${}^{10}$Be plus a neutron has been calculated and compared to experimental data~\cite{Pa03}. At next-to-leading order (NLO) one additional parameter associated with the asymptotic normalization coefficient of the 1/2$^+$ state can be adjusted to obtain a good description of the low-energy portion of this observable. The EFT's convergence pattern is as expected, given the nominal expansion parameter.

The resulting NLO prediction for the charge radius of the 1/2$^+$ state is consistent with the experiment~\cite{No09}. We also extracted the s-wave scattering length and effective range and the p-wave scattering volume that parametrize scattering of a neutron from ${}^{10}$Be. Lastly, we find ``universal" correlations between electromagnetic observables in halo nuclei with shallow 1/2$^+$ and 1/2$^-$ states.

\vspace*{-0.4cm} 
\setlength{\bibsep}{0.0em}
\begin{thebibliographynotitle}{99}
\vspace*{-0.2cm} 
\bibitem{HP}  D.~R.~Phillips and H.~Hammer,
arXiv:1103.1087. 

\bibitem{BHvK}
C.~Bertulani, H.~Hammer and U.~Van Kolck,
  Nucl.\ Phys.\ A  {\bf 712}  (2002) 37;
P.~Bedaque, H.~Hammer and U.~van Kolck,
  Phys.\ Lett.\  B {\bf 569} (2003) 159.
  
\bibitem{Nak1} T.~Nakamura {\it et al.}, Phys.\ Rev.\ Lett.\ {\bf 83} (1999)  1112.

\bibitem{Pa03}
R.~Palit {\it et al.}  
  Phys.\ Rev.\  C {\bf 68} (2003) 034318.

\bibitem{No09} W.~N\"orteshauser {\it et al.}, Phys.\ Rev.\ Lett. {\bf 102} 
(2009) 062503.
\end{thebibliographynotitle}

\newabstract 
\begin{center}
{\large\bf Electromagnetic Processes in Few-Nucleon Systems at Low Energy}\\[0.5cm]
L.\ Girlanda$^1$, S.\ Pastore$^2$, {\bf R.\ Schiavilla}$^3$, and M.\ Viviani$^4$  \\[0.3cm]
$^1$Dipartimento di Fisica, Universit\`a del Salento, I-73100 Lecce, Italy\\[0.3cm]
$^2$Physics Division, Argonne National Laboratory, IL 60439, USA\\[0.3cm]
$^3$Theory Center, Jefferson Lab, Newport News, VA 23606,\\
Department of Physics, Old Dominion University, Norfolk, VA 23529, USA\\[0.3cm]
$^4$INFN, Sezione di Pisa, I-56127 Pisa, Italy
\end{center}

Nuclear potentials and electromagnetic currents derived in
chiral effective field theory ($\chi$EFT) have been used recently
to study magnetic moments and radiative captures of thermal
neutrons on few-nucleon systems.  In the present talk, we
discuss the $\chi$EFT approach, specifically as it
pertains to the construction of the electromagnetic current
operator~\cite{ref1}, and provide a critical
review of results obtained so far for these processes~\cite{ref3}.
In particular, we show that the predicted cross sections for the
$n$-$d$ and $n$-$^3$He captures---both are induced by $M1$
transitions, which are strongly inhibited at the one-body level---are well reproduced
by theory.

We also address the issue of relativistic corrections
to chiral potentials~\cite{ref4}.  We sketch the methods used to
construct the most general, relativistically invariant, contact Lagrangian
at order $Q^2$, from which a complete, but non-minimal, set of
(contact) interaction terms is obtained.  In the non-relativistic
limit, these consist of 2 leading independent operator combinations of order $Q^0$,
accompanied by specific $Q^2$ corrections, and 7 sub-leading ones of order $Q^2$.
We show that this result also follows by enforcing the commutation relations among the
Poincar\'e group generators order by order in the low-energy expansion.  These
boost corrections should be taken into account in $\chi$EFT calculations
of nuclei with mass number $A>2$. 

\setlength{\bibsep}{0.0em}
\begin{thebibliographynotitle}{99}
\bibitem{ref1} S.\ Pastore, L.\ Girlanda, R.\ Schiavilla, and M.\ Viviani,
Phys.\ Rev.\ C {\bf 80}, 034004 (2009).
\bibitem{ref3} L.\ Girlanda, A.\ Kievsky, L.E.\ Marcucci, S.\ Pastore, R.\ Schiavilla, and M.\ Viviani,
Phys.\ Rev.\ Lett.\ {\bf 105}, 232502 (2010).
\bibitem{ref4} L.\ Girlanda, S.\ Pastore, R.\ Schiavilla, and M.\ Viviani,
Phys.\ Rev.\ C {\bf 81}, 034005 (2010).
\end{thebibliographynotitle}

\newabstract 
\begin{center}
{\large\bf The Power of $\chi$EFT: Connecting Nuclear structure and Weak Reactions}\\[0.5cm]
{\bf Doron Gazit}$^1$ \\[0.3cm]
$^1$Racah Institute of Physics, Hebrew University of Jerusalem, 91904, Jerusalem, Israel.\\[0.3cm]
\end{center}

The main success of $\chi$EFT is in supplying a perturbative microscopic Lagrangian applicable in the low energy nuclear regime. This Lagrangian is consistent with the symmetries of the fundamental theory, and in particular the approximate chiral symmetry.
This Lagrangian is used to derive not only a nuclear potential, but also Noether currents, exploiting the global SU(2)$_L\times$SU(2)$_R$ chiral symmetry. As this is the gauging of the electro-weak interaction, this current is the scattering operator in electro-weak processes. This connection has a very useful result: an inherent connection between strong observables, dictated by the potential, and electro-weak processes. In fact, the same low-energy constants (LECs) appear in both the force and the current. Another advantage is that, up to ${\mathcal O(Q})^3$, all LECs in the current, with the exception of one, can be fixed in pion-nucleon scattering. The extra LEC, coined $\hat{d}_R$, appears in a three-nucleon force term as well as in electro-weak two-body currents.

This connection has been used to fix the three-nucleon forces to high accuracy, facilitated by the accurate measurement of triton half-life \citetwo{1}{2}. In addition, using this relation corrects an unphysical trend of standard nuclear physics approach when applied to the beta-decay of $^6$He \cite{3}.

An extension of this approach to electro-weak processes in medium-heavy nuclei can be found in \cite{4}. There, normal ordering of the 2-body current is used to construct an effective single nucleon operator. This predicts a substantial effect on the rare $0\nu 2 \beta$ decay nuclear matrix element. In this application, $\chi$EFT supplies a consistent approach to extrapolate the current to the high momentum characteristic of these decay, from the low-energy regime of ordinary $\beta$- and $2\nu 2\beta$-decays.

One can conclude that this is a most effective, pragmatic, and already successful application of $\chi$EFT. This approach is and can be used to predict, without free parameters, electro-weak, as well as strong, observables in nuclei.

\setlength{\bibsep}{0.0em}
\begin{thebibliographynotitle}{99}
\bibitem{1} A. G{\aa}rdestig and D. R. Phillips, Phys. Rev. Lett., 98, 232301 (2006).
\bibitem{2} D. Gazit, S. Quaglioni, and P. Navratil, Phys. Rev. Lett. 103, 102502 (2009).
\bibitem{3} S. Vaintraub, N. Barnea, and D. Gazit, Phys. Rev. C 79, 065501 (2009).
\bibitem{4} J. Men\'{e}ndez, D. Gazit, and A. Schwenk, arXiv:1103.3622 (2011).
\end{thebibliographynotitle}

\newabstract 
\begin{center}
{\large\bf \hspace*{-0.2cm} 
 A high-accuracy calculation of the $\pi d$ scattering length}\\[0.5cm]
\hspace*{-0.1cm} {\bf V.~Baru}$^{1,2}$,  C.~Hanhart$^{1}$, M.\ Hoferichter$^{3}$, B.\ Kubis$^{3}$,
A.\ Nogga$^{1}$, D.\ R.\ Phillips$^{4}$\\[0.3cm]
$^{1}$ Forschungszentrum J\"ulich, Institut f\"ur Kernphysik,
J\"ulich Center for 
Hadron Physics  and Institute for Advanced Simulation,  D-52425 J\"ulich, Germany\\[0.1cm]
 $^{2}$ ITEP, 117218, B. Cheremushkinskaya 25, Moscow, Russia\\[0.1cm]
 $^{3}$ Helmholtz-Institut f\"{u}r Strahlen- und Kernphysik (Theorie), 
Bethe Center for Theoretical Physics, Universit\"at Bonn, D-53115 Bonn, Germany\\[0.1cm]
$^{4}$ Institute of Nuclear and Particle Physics and Department of Physics and Astronomy, 
         Ohio University, Athens, OH 45701, USA
\end{center}

For many years a combined analysis of pionic hydrogen and deuterium atoms has been known as a 
good tool to extract information on the isovector and especially on the isoscalar
$S$-wave $\pi N$ scattering length \cite{hadatoms}. However, given the smallness of the 
isoscalar scattering length, the analysis becomes useful  only 
if the pion-deuteron scattering length is controlled theoretically to a high accuracy 
comparable to the experimental precision. 
To achieve the required few-percent accuracy one needs  theoretical control over
all  isospin conserving 3-body $\pi NN\to \pi NN$ operators   up to one order before  the 
contribution of the dominant unknown $4N\pi\pi$ contact term.  This term appears 
at  next-to-next-to-leading order in Weinberg counting. The largest effect in the $\pi d$ scattering length
 stems from the double-scattering process, with  
the pion  scattered off two static (infinitely heavy) nucleons.
Among the most relevant corrections are the triple-scattering term \citetwo{beane}{liebig}, 
the effect of  nucleon recoil \cite{recoil}
related to the finite mass of the nucleon,  dispersive corrections due to the processes $\pi d\to NN \to \pi d$ and 
$\pi d\to \gamma NN \to \pi d$,  and the contribution which appears due to the treatment 
of the $\Delta(1232)$ resonance as an explicit degree of freedom \cite{dispdelta}. 
In addition, one needs to include isospin violating (IV) effects  
in both two-body ($\pi N$) \cite{Martin} and three-body operators. In Ref.~\cite{JOB} we 
accounted for  virtual-photon effects and  mass differences   in the  three-body operators.
We also included all the isospin-conserving effects listed above.
The resulting  combined analysis of $\pi H$ and $\pi D$ atoms yields: $ a^+ =(7.6\pm 3.1)\cdot 10^{-3} M_\pi^{-1}$ and $ a^-=(86.1\pm 0.9)\cdot 10^{-3} M_\pi^{-1}.$

\vspace*{-0.3cm}
\setlength{\bibsep}{0.0em}
\begin{thebibliographynotitle}{99}
\vspace*{-0.1cm}
\bibitem{hadatoms}
J.~Gasser, V.E.~Lyubovitskij, A.~Rusetsky,
Phys.\ Rept.\  {\bf 456} (2008) 167.

\bibitem{beane}
S.~R.~Beane et al., 
Nucl.\ Phys.\  A {\bf 720} (2003) 399.

\bibitem{liebig} 
S.~Liebig et al., 
arXiv:1003.3826 [nucl-th]. 

\bibitem{recoil}
V.~Baru et al., 
Phys.\ Lett.\  B {\bf 589} (2004)  118.

\bibitem{dispdelta} V.~Lensky et al., 
 Phys.\ Lett.\  B {\bf 648} (2007) 46;  Phys.Lett.B   {\bf 659} (2008) 184.

\bibitem{Martin}  M. Hoferichter et al., 
Phys.Lett.B {\bf 678} (2009) 65; Nucl. Phys.A {\bf 833} (2010)18.

\bibitem{JOB}  V. Baru et al., 
Phys. Lett. B  {\bf 694} (2011) 473. 
\end{thebibliographynotitle}

\newabstract 
\begin{center}
{\large\bf Hyperon-nucleon and hyperon-hyperon interactions in chiral effective
field theory}\\[0.5cm]
{\bf Johann Haidenbauer}\\[0.3cm]
Institut f\"ur Kernphysik and Institute for Advanced Simulation,
Forschungszentrum J\"ulich GmbH,
D-52425 J\"ulich, Germany\\[0.3cm]  
\end{center}

With regard to the nucleon-nucleon ($NN$) system a description of high
precision could be achieved within chiral effective field theory (EFT) 
\citetwo{Entem:2003ft}{Epe05}. Following the original suggestion of Steven 
Weinberg \cite{Wei90}, in these works the power counting is applied to the $NN$
potential rather than to the reaction amplitude. The latter is then obtained from
solving a regularized Lippmann-Schwinger equation for the derived interaction potential.
The $NN$ potential contains pion-exchanges and a series of contact interactions
with an increasing number of derivatives to parameterize the shorter ranged part
of the $NN$ force.

Recently, also baryon-baryon systems with strangeness
were investigated within the framework of chiral EFT by the group in J\"ulich
\citethree{Polinder:2006zh}{Polinder:2007mp}{Hai10}. 
In these works the same scheme as applied in Ref.~\cite{Epe05} to 
the $NN$ interaction is adopted. 
Specifically, the interactions in the $\Lambda N$ and $\Sigma N$
channels \cite{Polinder:2006zh} as well as those in the $S=-2$ sector
($\Lambda\Lambda$, $\Sigma\Sigma$, $\Lambda\Sigma$, $\Xi N$)
\cite{Polinder:2007mp} were considered at leading order (LO).
Moreover, predictions for the $S=-3$ and $-4$ baryon-baryon interactions 
were made \cite{Hai10}, invoking constraints from ${\rm SU(3)}$ flavor 
symmetry. To LO in the power counting the baryon-baryon
potentials involving strange baryons consist of four-baryon contact terms without
derivatives and of one-pseudoscalar-meson exchanges, analogous to the $NN$
potential \cite{Epe05}.
 
It turned out that already at LO the bulk properties of the $\Lambda N$ and $\Sigma N$
systems can be reasonably well accounted for. Furthermore, the EFT results are consistent 
with the rudimentary empirical information available in the $S=-2$ sector.  
Preliminary but still incomplete results for the $YN$ interaction to next-to-leading order
look very promising too \cite{Hai10a}. It will be interesting to see what can be achieved 
within a full calculation to NLO. 

\setlength{\bibsep}{0.0em}
\begin{thebibliographynotitle}{99}
\bibitem{Entem:2003ft}
D.~R. Entem, R.~Machleidt, Phys. Rev. C {\bf 68} (2003) 041001.

\bibitem{Epe05}
E.~Epelbaum, W.~Gl{\"o}ckle, U.-G. Mei{\ss}ner, Nucl. Phys. A {\bf 747} (2005) 362.

\bibitem{Wei90}
S.~Weinberg, Phys. Lett. B {\bf 251} (1990) 288;
Nucl. Phys. B {\bf 363} (1991) 3.

\bibitem{Polinder:2006zh}
  H.~Polinder, J.~Haidenbauer, U.-G.~Mei{\ss}ner,
  Nucl.\ Phys.\ A {\bf 779} (2006) 244.

\bibitem{Polinder:2007mp}
  H.~Polinder, J.~Haidenbauer, U.-G.~Mei{\ss}ner,
  Phys.\ Lett.\ B {\bf 653} (2007) 29.

\bibitem{Hai10}
J.~Haidenbauer, U.-G. Mei{\ss}ner,
  Phys.\ Lett.\ B \textbf{684} (2010) 275.

\bibitem{Hai10a} J.~Haidenbauer,
EPJ Web of Conferences \textbf{3} (2010) 01009.
\end{thebibliographynotitle}


\newabstract 
\begin{center}
{\large\bf Threshold resummation of heavy coloured-particle cross sections}\\[0.5cm]
Martin Beneke$^1$, {\bf Pietro Falgari}$^2$, Sebastian Klein$^1$ 
\\and  Christian Schwinn$^3$ \\[0.3cm]
$^1$Inst. f\"ur Theor. Teilchenphysik, RWTH Aachen University, Aachen, Germany\\[0.1cm]
$^2$Institute for Theor. Physics, Utrecht University, Utrecht, The Netherlands\\[0.1cm]
$^3$Albert Ludwigs Universit\"at Freiburg, Phys. Institut, 
Freiburg, Germany\\[0.5cm]
\end{center}

A precise theoretical description of the pair-production of heavy coloured particles
is hampered by 
quantum corrections to the partonic cross sections of the form $\alpha_s^n \ln^m \beta$ and $(\alpha_s/\beta)^k$, 
which arise from soft-gluon emission and Coulomb exchange, respectively. Near the
production threshold, defined by the limit $\beta \equiv \sqrt{1-4 M^2/\hat{s}} \rightarrow 0$, 
such terms are parametrically enhanced, possibly leading to a breakdown of perturbation theory, and should ideally be resummed
to all orders in the strong coupling $\alpha_s$. 
In this talk I discuss how resummation of soft and Coulomb corrections can be achieved through an effective-theory description of the pair-production
process, in which long-distance degrees of freedom with $q^2 \ll M^2$ are still dynamical, while the effect of \emph{hard} modes with $q \sim M$
is encoded in the effective couplings of the low-energy theory. 
In \cite{Beneke:2009rj} and  \cite{Beneke:2010da}, it has been shown that in such description the partonic cross section factorizes according to
\begin{equation}
\hat{\sigma} = \sum_S \sum_i H^S_i(\mu) \int d \omega \sum_{R_\alpha} J^S_{R_\alpha}(E-\omega/s) W^{R_\alpha}_i(\omega,\mu) \, ,
\end{equation} 
where $H_i^S$, $J^S_{R_\alpha}$ and $W^{R_\alpha}_i$ are determined, respectively, by hard, Coulomb and soft modes only. 
In this approach, large logarithms of $\beta$ are resummed to all orders by solving renormalization-group evolution equations for the 
functions $H_i^S$ and $W^{R_\alpha}_i$, while resummation of Coulomb singularities, contained in $J^S_{R_\alpha}$, is obtained via techniques developed in the context of PNRQCD.
The formalism has been recently applied to $t \bar{t}$ (Refs. \cite{Beneke:2009ye}, \cite{Beneke:2010fm})
and to squark-antisquark production (Ref. \cite{Beneke:2010da}), where resummation effects on the total cross section were found to be sizeable, and thus relevant for phenomenological
analysis at Tevatron and LHC.

\setlength{\bibsep}{0.0em}
\begin{thebibliographynotitle}{99}

\bibitem{Beneke:2009rj}
  M.~Beneke, P.~Falgari, C.~Schwinn,
  Nucl.\ Phys.\  {\bf B828 } (2010)  69-101.

\bibitem{Beneke:2010da}
  M.~Beneke, P.~Falgari, C.~Schwinn,
  Nucl.\ Phys.\  {\bf B842 } (2011) 414-474.

\bibitem{Beneke:2009ye}
  M.~Beneke, M.~Czakon, P.~Falgari, A.~Mitov, C.~Schwinn,
  Phys.\ Lett.\  {\bf B690 } (2010)  483-490.

\bibitem{Beneke:2010fm}
  M.~Beneke, P.~Falgari, S.~Klein, C.~Schwinn,
  Nucl.\ Phys.\ Proc.\ Suppl.\  {\bf 205-206 } (2010)  20-24.
\end{thebibliographynotitle}

\end{document}